\RequirePackage{ifpdf}
\ifpdf
\else
\PassOptionsToPackage{dvipdfmx}{graphicx}
\PassOptionsToPackage{dvipdfmx}{hyperref}
\fi

\documentclass[a4paper,11pt]{article}

\usepackage{jheppub} 

\usepackage[T1]{fontenc} 
\usepackage{appendix}
\usepackage{color}
\usepackage{mathrsfs}
\usepackage{amsmath,bm}
\usepackage{subfigure}
\usepackage{amssymb}
\usepackage{colortbl}
\usepackage{ulem}

\title{
    Observing black holes through superconductors
}

\author[a]{Youka Kaku\,}
\author[b]{\!, Keiju Murata\,}
\author[c]{\!, Jun Tsujimura\,}

\affiliation[a,c]{Department of Physics, Nagoya University, Chikusa, Nagoya 464-8602, Japan}
\affiliation[b]{Department of Physics, College of Humanities and Sciences, Nihon University, Sakurajosui, Tokyo 156-8550, Japan}

\emailAdd{kaku.yuka@e.mbox.nagoya-u.ac.jp}
\emailAdd{murata.keiju@nihon-u.ac.jp}
\emailAdd{tsujimura.jun@a.mbox.nagoya-u.ac.jp}

\abstract{
    We propose a way to observe the photon ring of the asymptotically anti-de Sitter black hole  dual to a superconductor on the two-dimensional sphere.
    We consider the electric current of the superconductor under the localized time-periodic external electromagnetic field.
    On the gravity side, the bulk Maxwell field is sent from the AdS boundary and then diffracted by the black hole.
    We construct the image of the black hole from the asymptotic data of the bulk Maxwell field that corresponds to the electric current on the field theory side.
    We decompose the electric current into the dissipative and non-dissipative parts and take the dissipative part for the imaging of the black hole.
    We investigate the effect of the charged scalar condensate on the image. 
    We obtain the bulk images that indicate the discontinuous change of the size of the photon ring.
}

\arxivnumber{}

\begin{document}
\maketitle
\flushbottom
\notoc

\section{Introduction}
\label{sec:Introduction}

    The AdS/CFT correspondence~\cite{Maldacena:1997re,Gubser:1998bc,Witten:1998qj} is a duality between strongly coupled quantum field theories (QFTs) and classical gravity in AdS spacetime. 
    In recent years, it has been conjectured that the duality can  describe  realistic systems such as condensed matter physics~\cite{Hartnoll:2009sz,Herzog:2009xv,McGreevy:2009xe,Horowitz:2010gk,Sachdev:2010ch}. 
    In particular, we can analyze the thermal states of QFTs  by the black hole physics in AdS/CFT. 
    How can we directly test the existence of the dual black hole for a given QFT? We will address this problem in this paper.

    In astronomy, the Event Horizon Telescope~\cite{Akiyama:2019cqa} succeeded in constructing the first image of the supermassive black hole in M87. 
    They observed the photon sphere of the black hole, and it is one of the most direct observations of the black hole in our real space. 
    On the other hand, there is a proposal that a similar observation of the black hole through AdS/CFT \cite{Hashimoto:2018okj,Hashimoto:2019jmw}.
    They considered QFT on $R_t\times S^2$ and applied a localized time-periodic external source in $S^2$. 
    In the dual gravitational picture, the external source is mapped to the boundary condition of the bulk field. 
    Because of the time-dependant boundary condition, a  wave is sent from the AdS boundary and propagates through the black hole spacetime. 
    The wave is diffracted by the black hole and, eventually, arrives at other points on the AdS boundary. 
    From the asymptotic data of the bulk field, which corresponds to the response function on the QFT side, 
    they constructed the image of the AdS black hole.
    The formula for converting the response function to the image of the black hole has also been obtained in Refs.\cite{Hashimoto:2018okj,Hashimoto:2019jmw}.
    In this paper, we apply this idea  to the model of the holographic superconductor.

    In this paper, we consider imaging of the black hole dual to a superconductor.
    A model of the holographic superconductor on $S^2$ is composed of a Maxwell field and charged scalar field in a fixed spherical AdS black hole background~\cite{Gubser:2008px,Hartnoll:2008vx,Hartnoll:2008kx}.
    (We only focus on the probe limit of the holographic superconductor.) 
    When the Hawking temperature is smaller than a critical temperature $T<T_c$, the AdS black hole becomes unstable against charged scalar field perturbation. 
    Resultantly, the $U(1)$-gauge symmetry is spontaneously broken, and the black hole with the charged scalar field hair is to realize as a stable configuration~\cite{Murata:2010dx,Kanno:2010pq,Bhaseen:2012gg,Bosch:2016vcp}.
    We can identify this phase transition  as the superconducting phase transition.
    We consider a laser applied to one point of the superconductor on $S^2$.
    From the dual gravitational point of view, the external electromagnetic field is regarded as the boundary condition of the bulk Maxwell field.
    We construct the image of the black hole through the response function of the electromagnetic field, i.e., the electric current.
    The electric current can be decomposed into the dissipative and non-dissipative parts (i.e., modes with phase difference $\phi=0, \pi/2$ with respect to the applied external field.).
    We will see that the information of the photon ring is mainly encoded in the dissipative part.
    We will also investigate how the superconducting phase transition affects the image of the AdS black hole.

    This paper is structured as follows.  
    In the next section, we review the previous work~\cite{Hashimoto:2018okj,Hashimoto:2019jmw} in which the image formation of the AdS black hole.
    In section \ref{sec:Setup}, we introduce the holographic model for the superconductor on $S^2$. We explicitly construct the gravitational solution in the superconducting phase.
    In section \ref{sec:perturbation}, we consider the linear perturbation on the holographic superconductor. It is decomposed into vector and scalar modes. 
    In section \ref{sec:imaging_sc}, we show images of black holes dual to the superconductor. We estimate the radius of the photon ring in the image and found that it changes discontinuously for the vector mode.
    The final section is devoted to the conclusion.

\section{Bulk imaging through AdS/CFT}
\label{sec:Review}

    We will review the imaging black hole through AdS/CFT correspondence \cite{Hashimoto:2018okj, Hashimoto:2019jmw}. For simplicity, let us consider a minimal coupled scalar field $\Psi$ in Schwarzchild-AdS$_4$ spacetime (Sch-AdS$_4$), 
    \begin{equation} \label{eq:SAdS metric}
        ds^2 = -F(r)dt^2 + \frac{dr^2}{F(r)} + r^2 (d\theta^2 + \sin^2\theta d\varphi^2)\ ,
    \end{equation}
    \begin{equation} \label{eq:Lapse function}
        F(r) = 1 + \frac{r^2}{L^2} - \dfrac{r_h}{r}\left( 1 + \frac{r_h^2}{L^2} \right)\ ,
    \end{equation}
    where $r_h$ is the radius of black hole horizon and $L$ is the AdS radius.
    We define the tortoise coordinate $r_*$ as
    \begin{equation}
        r_*=\int_\infty^r \frac{dr}{F(r)}\ .
    \end{equation}
    The scalar field $\Psi$ obeys the following Klein-Gordon equation.
    \begin{equation} 
        -\frac{1}{F} \partial_t^2 \Psi + F \partial_r^2 \Psi + \frac{(r^2 F)'}{r^2} \partial_r \Psi + \frac{1}{r^2} D^2\Psi = 0\ ,
    \end{equation}
    where the prime denotes $r$-derivative and $D^2$ is the scalar Laplacian on unit $S^2$.

    In the vicinity of the AdS boundary, the scalar field $\Psi$ in this system  behaves 
    \begin{equation} \label{eq:Klein Gordon}
        \Psi(t,r,\theta,\varphi) = S_{\mathcal{O}}(t,\theta,\varphi) -\frac{1}{2r^2}(\partial^2_t-D^2)S_{\mathcal{O}}(t,\theta,\varphi) + \frac{\langle \mathcal{O}(t,\theta,\varphi) \rangle}{r^3} + O(r^{-4})\ .
    \end{equation}
    According to the AdS/CFT dictionary, $S_{\mathcal{O}}$ and $\langle \mathcal{O}\rangle$ are two independent functions, and we can regard them as the external scalar source and its response function in the dual CFT, respectively. Note that when we deal with other fields like a vector field, we should be careful which coefficient corresponds to the dual CFT value we want. We will mention  the holographic superconductor case in Section \ref{sec:perturbation} and Appendix \ref{appendix_response}.

\begin{figure}[ht]
        \centering
		\includegraphics[width=80mm]{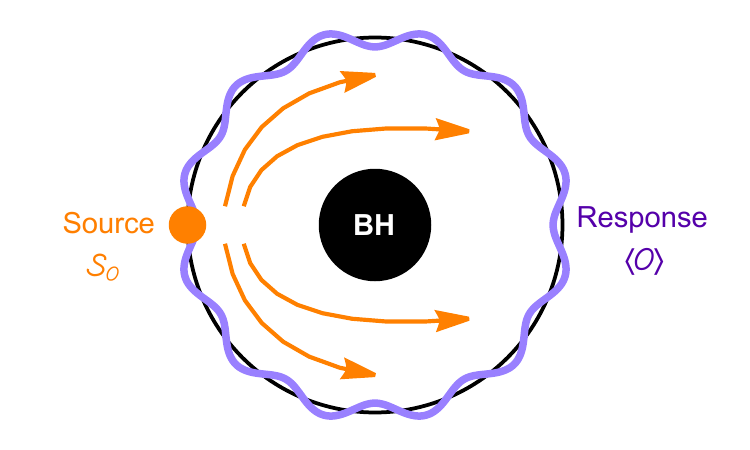}
        \caption{Source and response}
        \label{fig:source_response}
    \end{figure}

    Let $S_{\mathcal{O}}$ be the following axisymmetric and monochromatically oscillating Gaussian source localized at $\theta=\pi$ as
    \begin{equation}
        S_{\mathcal{O}}(t,\theta,\varphi) = e^{-i \omega t} g(\theta)\ ,
        \label{JO}
    \end{equation}
    \begin{equation}
    \label{eq:Gaussian_source}
    g(\theta) = \frac{1}{2\pi \sigma^2} \exp{ \left[-\frac{(\pi-\theta)^2}{2\sigma^2} \right]}\ .
    \end{equation}
    For $\sigma\ll 1$, the Gaussian function is decomposed into the scalar spherical harmonics as 
    \begin{equation}
    g(\theta) \simeq  \sum_{l=0}^\infty c_l Y_{l0}\ ,\quad  c_l\equiv 
    (-1)^l \sqrt{\frac{l+1/2}{2\pi}}\exp\left[-\frac{1}{2}(l+1/2)^2 \sigma^2\right]\ .
    \label{eq:vector_mode_cl}
    \end{equation}
    Eq.~(\ref{JO}) gives the normalization condition of $\Psi$ at the AdS boundary. We also impose the in-going boundary condition at the horizon of the Sch-AdS$_{4}$. Then, we have the unique solution of \eqref{eq:Klein Gordon} and, in particular, the response function $\langle \mathcal{O}(t,\theta,\varphi) \rangle$ as the coefficient of $r^{-3}$ term. Schematically, the source $S_{\mathcal{O}}$ at the AdS boundary excites the scalar field, propagates in the bulk, and reaches another point at the AdS boundary as depicted in Fig.~\ref{fig:source_response}. We can get the picture of a bulk object like a black hole by imaging the response function through the lens.

\begin{figure}[ht]
        \centering
        \includegraphics[width=130mm]{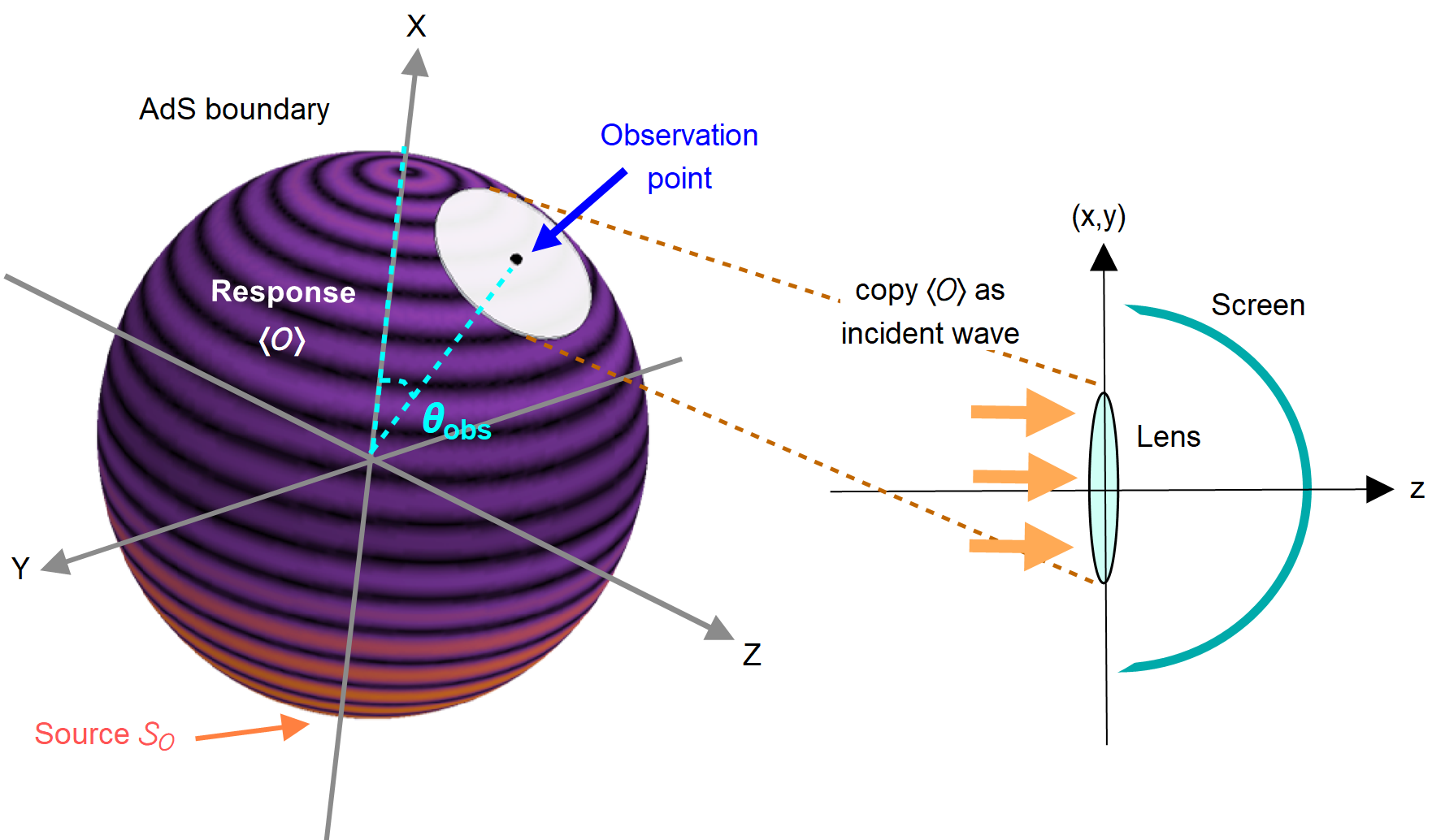}
        \caption{Schematic diagram of a setup. 
        We will project the response function onto the lens in the virtual optical system and build its image on the hemispherical screen.}
        \label{fig:AdS_boundary}
    \end{figure}

    Next, we introduce a virtual optical system in flat space and create the image of the response function living on the AdS boundary, as if we observe the bulk figure from the boundary. (Note that this optical system is in neither the bulk spacetime nor the boudary $S^2$.) In particular, we will apply the Fourier-Fresnel transformation to the response function on a small patch of the AdS boundary. (See also \cite{Nambu2013,Kanai2013,Nambu2016}.)
    We depicted a schematic picture of our setup in Fig.~\ref{fig:AdS_boundary}. The picture on the left side shows the AdS spacetime with the source at the south pole and the response $\langle \mathcal{O}(t,\theta,\varphi) \rangle$ on the AdS boundary. We set the observation point at $\theta=\theta_{\rm{obs}}$ on the boundary. The picture on the right side shows the virtual optical system in the 3-dimensional flat space $(x,y,z)$ with a thin convex lens and a hemispherical screen. We will read off the response on a small patch around the observation point on the AdS boundary, copy the response function to the virtual optical system as the incident wave on the lens, and build its image on the screen.
    
\begin{figure}[ht]
        \centering
        \includegraphics[width=70mm]{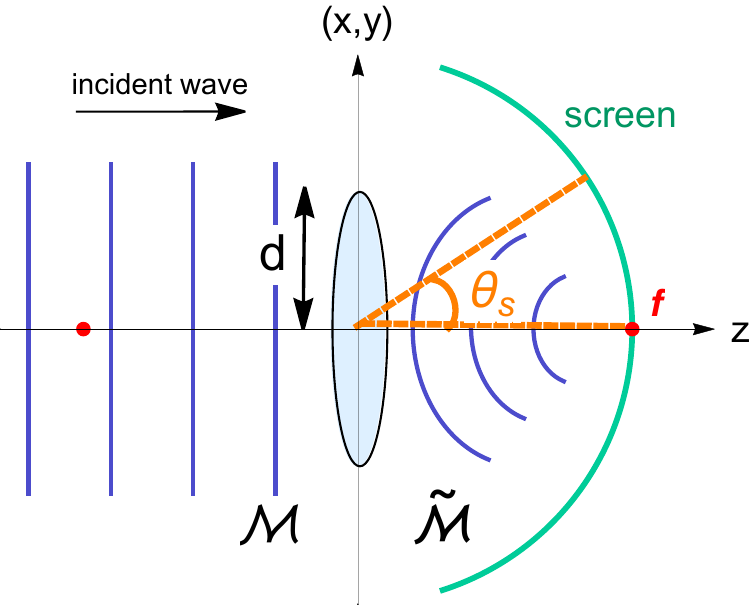}
        \caption{A virtual wave optical system, where the lens is located at $\{(x_L,y_L)\equiv(x,y,0)|x^2+y^2\leq d^2\}$, and the screen is located at $\{(x_s,y_s,z_s)\equiv(x,y,z)|x^2+y^2+z^2=f^2, z\geq0\}$.}
        \label{fig:wave_optics}
    \end{figure}

    Let us consider a virtual optical system as depicted in Fig.~\ref{fig:wave_optics}. The thin lens is located at $\{(x_L,y_L)\equiv(x,y,0)|x^2+y^2\leq d^2\}$, and the screen is located at $\{(x_s,y_s,z_s)\equiv(x,y,z)|x^2+y^2+z^2=f^2, z\geq0\}$, where $f$ is the focal length \cite{Nambu2013,Kanai2013,Nambu2016}.
    The incident plane wave comes from $z<0$ and the lens emits spherical waves to $z>0$. Let $\mathcal{M}(\vec{x})$ and $\tilde{\mathcal{M}}(\vec{x})$ be the wave function of a monochromatic incident wave and diffracted wave with frequency $\omega$, respectively. According to the wave optics \cite{hecht2012optics}, the conversion of $\mathcal{M}$ to $\tilde{\mathcal{M}}$ is given by 
    \begin{equation}
        \tilde{\mathcal{M}}(\vec{x})=e^{-i \omega \frac{|\vec{x}|^2}{2 f}}\mathcal{M}(\vec{x})\ .
    \end{equation}
   The wave function on the screen $\mathcal{I}(\vec{x_s})$ is expressed as the sum of wave functions which are emitted from every point on the lens:
    \begin{equation}
        \mathcal{I}(\vec{x_s})
        =
        \int_{|\vec{x}| \leq d} d^2 x \tilde{\mathcal{M}}(\vec{x}) e^{i \omega |\vec{x_s}-\vec{x}|}\\
        \simeq
        e^{i \omega f}
        \int_{|\vec{x}| \leq d} d^2 x \mathcal{M} (\vec{x}) e^{i \omega \vec{x}\cdot\frac{\vec{x_s}}{f}}
    \end{equation}
    Here, we assumed $d \ll f$ for the second equality.
    Hence the image constructed by the incident wave $\mathcal{M}(\vec{x})$ is expressed as
    \begin{equation}
        \label{eq:lens_formula}
        |\mathcal{I}(\vec{x_s})|^2
        \simeq
        \left|
        \int_{|\vec{x}| \leq d} d^2 x \mathcal{M} (\vec{x}) e^{i \omega \vec{x}\cdot\frac{\vec{x_s}}{f}}
        \right|^2\ .
    \end{equation}

    To apply \eqref{eq:lens_formula} to the response, we need to transform the $S^2$ coordinate on the AdS boundary $(\theta,\phi)$ to a Cartesian coordinate on the lens $(x_L,y_L)$.  
    First, the Cartesian coordinate on the AdS boundary (X,Y,Z) is given as
    \begin{equation}
        X(\theta,\phi)=\sin{\theta} \cos{\phi}\ ,
        \ \ 
        Y(\theta,\phi)=\sin{\theta} \sin{\phi}\ ,
        \ \ 
        Z(\theta,\phi)=\cos{\theta}\ .
    \end{equation}
    Next, rotate it around the $Y$-axis so that the observation point $\theta=\theta_{\rm{obs}}$ comes to the north pole. The rotated Cartesian coordinate $(X',Y',Z')$ is
    \begin{align}
        &X'(\theta,\phi,\theta_{\rm{obs}})
        =\cos{\theta_{\rm{obs}}}~X(\theta,\phi)-\sin{\theta_{\rm{obs}}}~Z(\theta,\phi)\ ,\\
        &Y'(\theta,\phi)
        =Y(\theta,\phi)\ ,\\
        &Z'(\theta,\phi,\theta_{\rm{obs}})
        =\sin{\theta_{\rm{obs}}}~X(\theta,\phi)+\cos{\theta_{\rm{obs}}}~Z(\theta,\phi)\ .
    \end{align}
    Then, the rotated $S^2$ coordinate $(\theta', \phi')$ on the AdS boundary is
    \begin{align}
        \theta'(\theta,\phi,\theta_{\rm{obs}})
        =\arccos{Z'\left(\theta,\phi,\theta_{\rm{obs}}\right)}\ ,
        \ \ 
        \phi'(\theta,\phi,\theta_{\rm{obs}})
        =\arctan{\left(\frac{Y'(\theta,\phi)}{X'(\theta,\phi,\theta_{\rm{obs}})}\right)}\ .
    \end{align}
    Finally, we take a sufficiently small patch around $\theta'=0$ ($\theta=\theta_{\rm{obs}}$) on the AdS boundary, and regard $(\theta',\phi')$ as the radial coordinate and the angular coordinate on the lens, respectively. Therefore, we obtain the following transformation formula of $(\theta,\phi) \rightarrow (x_L,y_L)$.
    \begin{align}
        x_L(\theta,\phi,\theta_{\rm{obs}})
        &=\theta'(\theta,\phi,\theta_{\rm{obs}})\cos{\left(\phi'(\theta,\phi,\theta_{\rm{obs}})\right)}\ ,\\
        y_L(\theta,\phi,\theta_{\rm{obs}})
        &=\theta'(\theta,\phi,\theta_{\rm{obs}})\sin{\left(\phi'(\theta,\phi,\theta_{\rm{obs}})\right)}\ .
    \end{align}
    We can construct the bulk image viewed at $\theta= \theta_{obs}$ by evaluating \eqref{eq:lens_formula} substituted the response function on this coordinate system $(x_L, y_L)$.

\begin{figure}[htbp]
	\begin{tabular}{cc}
	\begin{minipage}[t]{.49\textwidth}
        \centering
        \includegraphics[width=70mm]{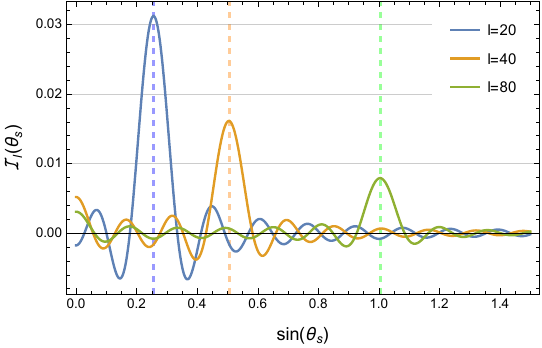}
        \caption{The behaviour of $\mathcal{I}{(\vec{x_s})}$ for $d=0.5, \omega=80$, and $l=20,40,80$. The dashed lines represents $\theta_s=\arcsin\sqrt{\lambda_l}/\omega$.}
        \label{fig:background_sols}
	\end{minipage}
	\begin{minipage}[t]{.49\textwidth}
        \centering
        \includegraphics[width=65mm]{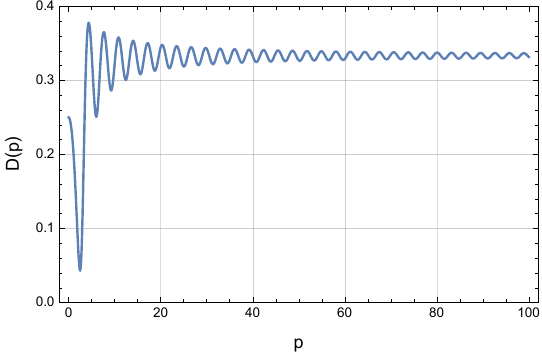}
        \caption{The behavior of $\mathrm{D}(p).$}
        \label{fig:D(p)}
	\end{minipage}
	\end{tabular}
\end{figure}

    Before going to the next section, let us see the image of $l$ mode spherical wave $\mathcal{M} (\vec{x}) = Y_{l0}(\theta) \simeq P_l(\cos\theta)$, where $P_l$ is the Legendre function. It may be useful to understand the behavior of the image.
    We can perform the integration of \eqref{eq:lens_formula} analytically when $|\vec{x}| \le d \ll f$. In this case, note that $P_l(\cos \theta) \simeq J_0({\sqrt{\lambda_l}\,\theta})$, where $\ \lambda_l = l(l+1)$ and $J_n(x)$ is the Bessel function of the first kind. Then,
    \begin{align}
        \mathcal{I}_l(\vec{x_s}) 
        &= \int^d_0 d\theta\int^{2\pi}_0 d\varphi\, 
        \theta J_0(\sqrt{\lambda_l}\,\theta) e^{i \omega \theta\cdot\sin\theta_s \cos{(\varphi-\varphi_s)}}\nonumber  \\
        \label{eq:analiticalFT} 
        &=
        d^2 \Delta(d\sqrt{\lambda_l},d\omega \sin\theta_s).
    \end{align}
    Here, $(\theta_s,\varphi_s)$ denote angular coordinates of the hemisphere screen as we depicted in Fig.~\ref{fig:wave_optics}, and
\begin{align}
	\Delta(p,q)\equiv\frac{2\pi}{p^2-q^2}\left[
	    p J_1(p) J_0(q)-q J_1(q) J_0(p)
	    \right].
\end{align}
    Notice that the image amplitude of $l$ component $|\mathcal{I}_l(\vec{x_s})|^2$ takes some large value at $\theta_s=\arcsin\sqrt{\frac{\lambda_l}{\omega^2}}$ since there exists $\lambda_l-\omega^2\sin \theta_s^2$ factor in a denominator of \eqref{eq:analiticalFT}. (Although, its value is finite: $\pi d^2 \left[J_0\left(d\omega\sin\theta_s\right)^2 + J_1\left(d\omega\sin\theta_s\right)^2\right]$.) We also depicted the typical behavior of $\mathcal{I}_l{(\vec{x_s})}$ in Fig.~\ref{fig:background_sols}. Roughly speaking, the $l$ mode component wave yields a ring with a radius $\sin\theta_s=\sqrt{{\frac{\lambda_l}{\omega^2}}}$ by the Fourier-Fresnel transformation in \eqref{eq:lens_formula}. 
    Hence, when a superposed wave packet of several $l$ mode spherical waves comes into the lens, we see the ring image of a dominant coefficient mode.
    
	Although the Fresnel formula \eqref{eq:analiticalFT} is valid as long as $d \ll f$, the frequency $\omega$ of the propagating field should be high enough to build high-resolution images. The resolution of image, or the variance of image amplitude is naively determined by the second-order derivative of image amplitude:
	\begin{equation}
	    \left[-\frac{1}{\mathcal{I}_l(\vec{x_s})}\frac{\partial^2\mathcal{I}_l(\vec{x_s})}{\partial(\sin\theta_s)^2}\right]_{\sin\theta_s=\sqrt{\lambda_l/\omega^2}}
	    =d^2\omega^2\left[-\frac{1}{\Delta(p,q)}\frac{\partial^2\Delta(p,q)}{\partial q^2}\right]_{q=p}
	    =d^2 \omega^2 \mathrm{D}(p),
	\end{equation}
    where $p=d\sqrt{\lambda_l}, q=d\omega\sin\theta_s$, and
	\begin{equation}
	    \mathrm{D}(p)
	    \equiv\left[-\frac{1}{\Delta(p,q)}\frac{\partial^2\Delta(p,q)}{\partial q^2}\right]_{q=p}
	    =\frac{p^2 J_0(p)^2+2 p J_0(p)J_1(p)+(p^2-5)J_1(p)^2}{3 p^2 \left(J_0(p)^2+J_1(p)^2\right)}.
	\end{equation}
	We depicted the behaviour of $\mathrm{D}(p)$ in Fig.~\ref{fig:D(p)}, which shows $\mathrm{D}(p)\sim \mathcal{O}(1)$. Then, the condition to get high-resolution image can be replaced in context of $d$ and $\omega$ as follows.
	\begin{equation}
	    \left[-\frac{1}{\mathcal{I}_l(\vec{x_s})}\frac{\partial^2\mathcal{I}_l(\vec{x_s})}{\partial(\sin\theta_s)^2}\right]_{\sin\theta_s=\sqrt{\lambda_l/\omega^2}}\gg 1
	    \ \ \leftrightarrow\ \ 
	    d^2 \omega^2 \gg 1
	\end{equation}
	Hence, we should set the parameters to satisfy $\omega \gg 1/d$ to obtain high-resolution images.
    That is, if the wavelength of the incident wave is small enough compared to the lens radius, we get a high-resolution image with less wave effect.

    For an intuitive understanding of the image of the  black hole, it is convenient to consider the null geodesic in Sch-AdS$_4$. 
    (See also Ref.\cite{Hashimoto:2018okj} for detailed analysis.)
    We assume that the orbital plane of the null geodesic is in the equatorial plane $\theta=\pi/2$.
    Then, from the geodesic equation, we have
    \begin{equation}
        \dot{r}^2=\omega^2-\ell^2 v(r)\ ,\quad v(r)\equiv \frac{F(r)}{r^2}\ ,
        \label{geodesic}
    \end{equation}
    where $\omega=F(r)\dot{t}$ and $\ell=r^2\dot{\varphi}$ are the conserved energy and angular momentum, respectively. 
    The dot denotes the derivative by an affine parameter.
    The effective potential $v(r)$ has a maximal value
    \begin{equation}
        v_\text{max}=\frac{(1+3r_h^2)^2(4+3r_h^2)}{27r_h^2(1+r_h^2)^2}\ ,
        \label{vmax}
    \end{equation}
    at $r=r_\textrm{max}=3(1+r_h^2)r_h/2$. 
    At the photon sphere $r=r_\textrm{max}$, there is an unstable circular orbit satisfying $\dot{r}=0$.
    The angular momentum per unit energy of such a null geodesic is given by
    \begin{equation}
        \left(\frac{\ell}{\omega}\right)_\textrm{photon sphere} = \frac{1}{\sqrt{v_\textrm{max}}}\ .
        \label{lphoton}
    \end{equation}
    In the wave picture, the null geodesic with the angular momentum $\ell$ is composed of the spherical harmonics $Y_{l0}$ with $\ell\leq l \leq \ell +\Delta \ell$ ($\Delta \ell \ll \ell$).
    Therefore, the radius of the photon ring predicted by the geodesic analysis is 
    \begin{equation}
        \sin \theta_s = \frac{1}{\sqrt{v_\textrm{max}}}\ .
        \label{Estnradius}
    \end{equation}
	As raising a horizon radius $r_h$, $v_{\rm{max}}$ becomes smaller and the photon ring radius $\theta_s$ becomes larger.

    In the following of this paper, we will make the image of linear perturbative electromagnetic field on the background charged scalar and electromagnetic field in background Sch-AdS$_4$ spacetime. 
    We will show what asymptotic coefficients correspond to the source and its response as required. Then, we construct the observable from the response and built its image through \eqref{eq:lens_formula}.

\section{Holographic superconductor on \texorpdfstring{$S^2$}{Lg}}
\label{sec:Setup}

    We consider the s-wave holographic superconductor without the back reaction from the gravity \cite{Hartnoll:2008vx}. Consider the following  Einstein-Maxwell-charged scalar system, of which Lagrangian density is 
    \begin{equation}
        \mathcal{L} = \dfrac{1}{16\pi G}\left( R+\dfrac{6}{L^2} \right) -\frac{1}{4} F_{\mu \nu} F^{\mu \nu} 
        - \mathcal{D}_\mu \Psi (\mathcal{D}^\mu \Psi)^* + \frac{2|\Psi|^2}{L^2}\ ,
    \end{equation}
    where $G, R, L, F_{\mu \nu}, \mathcal{D}_\mu, \Psi$ are the gravitational constant, the Ricci scalar, the AdS radius, the field strength, the covariant derivative with respect to the background metric and the $U(1)$ gauge field, and the charged scalar field, respectively. 
    In our actual calculations, we take the unit of $L=1$.
    We will consider the scalar field and the $U(1)$ gauge field as the probe fields. This is achieved by taking $G \to 0$. (See Ref.\cite{Basu:2010uz} for the back reacted case.) Then, we can choose the Sch-AdS$_4$~(\ref{eq:SAdS metric}) as the background spacetime solution.\footnote{
    We will consider both of the small ($r_h<L$) and large ($r_h\geq L$) black hole branches. 
    The CFT dual of the small black hole has been studied in several literature~\cite{Hollowood:2006xb,Asplund:2008xd,Hanada:2016pwv,Yaffe:2017axl,Marolf:2018ldl,Choi:2021lbk}. 
    In the CFT dual of the small black hole, we would be able to see the its image.
    }
    The charged scalar field $\Psi$ and the $U(1)$ gauge field $A_{\mu}$ follows the equations below.
    \begin{equation} \label{eq:EoM_scalar}
        \mathcal{D}_{\mu}\mathcal{D}^{\mu}\Psi+\frac{2\Psi}{L^2}=0\ ,
    \end{equation}
    \begin{equation} \label{eq:EoM_gauge}
        \nabla_{\nu}F^{\nu\mu} = J^\mu
        ,\ \ \ J^\mu\equiv i\left( \Psi^* \mathcal{D}^{\mu} \Psi-\Psi \left( \mathcal{D}^{\mu} \Psi \right)^* \right)\ .
    \end{equation}
    In this section, we will solve these equations of motion.

\subsection{Normal and superconducting phases of holographic superconductor}

    Equations of motion~(\ref{eq:EoM_gauge}) have a solution
    \begin{equation}
        \Psi=0\ ,\quad \Phi=\rho \left(\frac{1}{r_h}-\frac{1}{r}\right)\ .
        \label{normal}
    \end{equation}
    Here, we set $\Phi(r_h)=0$ as the $U(1)$-gauge condition. The constant $\rho$ is the $U(1)$-charge.
    In this solution, the charged scalar field is trivial and $U(1)$-gauge symmetry in the bulk is preserved.
    This solution has been identified as the normal phase of the superconductor~\cite{Hartnoll:2008vx}.
    As we increase the charge $\rho$ for a fixed horizon radius $r_h$, 
    the normal phase solution~(\ref{normal}) becomes unstable against the charged scalar field perturbation~\cite{Gubser:2008px}.
    At the onset of the instability, there is a normal mode of the charged scalar field perturbation.
    We can extend the normal mode to a nonlinear regime.
    Such a solution has the charged scalar hair and the $U(1)$-symmetry is spontaneously broken. 
    This solution has been identified as the superconducting phase of the superconductor.

\begin{figure}[htbp]
	\begin{tabular}{cc}
	\begin{minipage}{.45\textwidth}
        \centering
        \includegraphics[width=65mm]{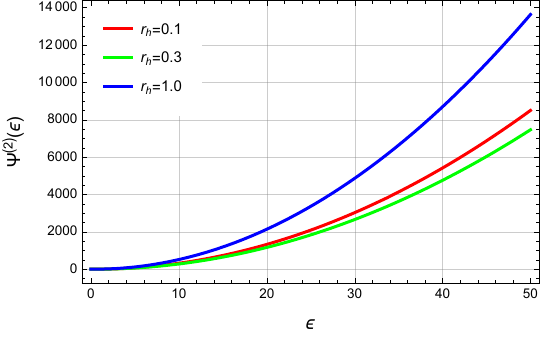}
        \caption{$\Psi^{(2)}(\epsilon)$ for $r_h=0.1, 0.3, 1.0$}
        \label{fig:psi2_epsilon}
	\end{minipage}
	\hspace{5mm}
	\begin{minipage}{.45\textwidth}
        \centering
        \includegraphics[width=65mm]{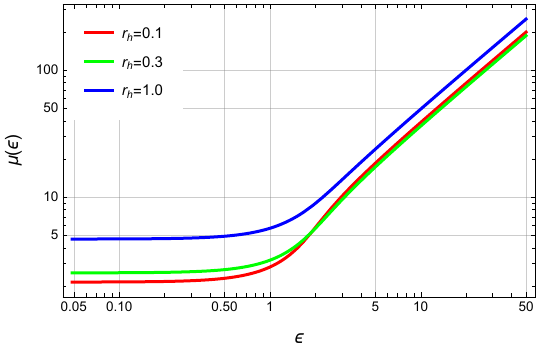}
        \caption{$\mu(\epsilon)$ for $r_h=0.1, 0.3, 1.0$}
        \label{fig:mu_epsilon}
	\end{minipage}
	\end{tabular}
\end{figure}

\begin{figure}[htbp]
	\begin{tabular}{cc}
	\begin{minipage}{.45\textwidth}
        \centering
        \includegraphics[width=65mm]{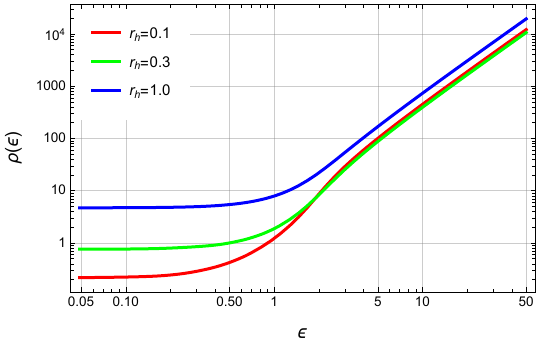}
        \caption{$\rho(\epsilon)$ for $r_h=0.1, 0.3, 1.0$}
        \label{fig:rho_epsilon}
	\end{minipage}
	\hspace{5mm}
	\begin{minipage}{.45\textwidth}
        \centering
        \includegraphics[width=65mm]{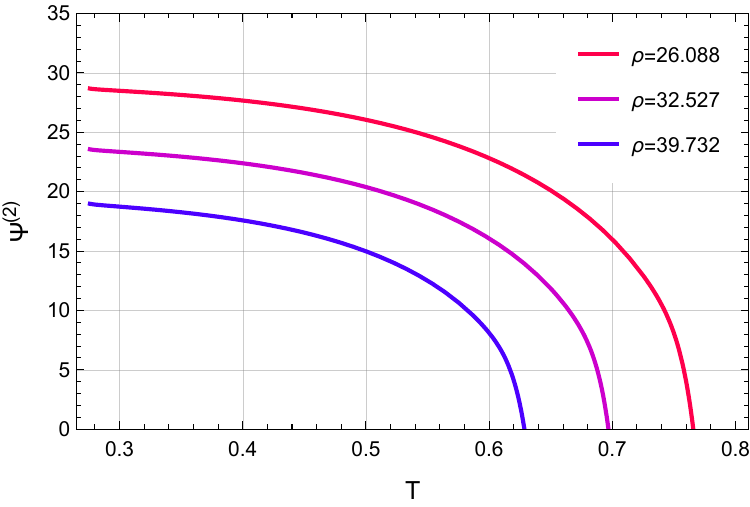}
        \caption{The condensation of $\Psi^{(2)}$ with respect to the Hawking temperature $T(r_h)$, where $\rho$ is set to be constant.}
        \label{fig:condensation}
	\end{minipage}
	\end{tabular}
\end{figure}

    Let us explicitly construct the solution in the superconducting phase.
    Under the spherically symmetric ansats, $\Phi=\Phi(r)$ and $A_\mu = \Phi(r)\delta^t_{\ \mu}$, equations of motion~\eqref{eq:EoM_scalar} become
    \begin{equation} \label{eq:background_scalar}
        \Psi(r)'' + \left( \frac{F(r)'}{F(r)}+\frac{2}{r} \right)\Psi(r)' + \frac{\Phi(r)^2}{F(r)^2}\Psi(r) + \frac{2}{L^2 F(r)}\Psi(r) = 0\ ,
    \end{equation}
    \begin{equation} \label{eq:background_gauge}
        \Phi(r)'' + \frac{2}{r}\Phi(r)' -\frac{2 \Psi(r)^2}{F(r)}\Phi(r)=0\ .
    \end{equation}
    We again set $\Phi(r_h)=0$ as the $U(1)$-gauge condition. We can also assume that $\Psi(r)$ is a real valued function by choosing its appropriate phase.
    Then, solving the above equations near the horizon, we obtain the regularity condition of the charged scalar field as $\Psi'(r_h)=-2\epsilon/(3r_h)$,
    where we define the horizon value of the scalar field as
    \begin{equation}
        \epsilon \equiv \Psi(r_h)\ ,\quad q\equiv r_h^2 \Phi'(r_h)
    \end{equation}
    Thus, the regular solution at the horizon is parameterized by three parameters, $(r_h,\epsilon,q)$ in the unit of $L=1$.
    On the other hand, the asymptotic behaviours of these fields near the AdS boundary are
    \begin{equation}
        \Psi(r \to \infty) = \frac{\Psi^{(1)}}{r} + \frac{\Psi^{(2)}}{r^2} + \cdots\ ,
        \ \ \ 
        \Phi(r \to \infty) = \mu - \frac{\rho}{r} + \cdots\ ,
    \end{equation}
    where $\Psi^{(1)},\Psi^{(2)},\mu,\rho$ are constants with respect to $r$. 
    As a new condition, we impose $\Psi^{(1)}=0$. This condition determines the value of $q$ for fixed $r_h$ and $\epsilon$.
    Therefore, the solutions satisfying the boundary conditions at the horizon and the infinity are  specified by the two parameters $(r_h,\epsilon)$.
    Then, for a fixed horizon radius $r_h$, we can regard the other constants as functions of $\epsilon$ like $q(\epsilon), \Psi^{(2)}(\epsilon), \mu(\epsilon), \rho(\epsilon)$. 
    Figs.\ref{fig:psi2_epsilon}, \ref{fig:mu_epsilon} and \ref{fig:rho_epsilon} show the functional profiles of QFT values $\Psi^{(2)}(\epsilon), \mu(\epsilon)$ and $\rho(\epsilon)$, respectively.

    In the boundary QFT point of view, it is convenient to specify the solutions by $\mu$ or $\Psi^{(2)}$ since we can regard them as the chemical potential and the order parameter of the corresponding QFT respectively, according to the GKP-Witten relation. However, we will choose $\epsilon$ as the parameter specifying the solution in this paper, because we focus on phenomena on the gravity side. We can easily map $(r_h,\epsilon)$ to quantities of QFT using Figs. \ref{fig:psi2_epsilon}, \ref{fig:mu_epsilon} and \ref{fig:rho_epsilon}.

    In Fig.~\ref{fig:condensation}, we show the behaviour of $\Psi^{(2)}$ with respect to the Hawking temperature $T(r_h)$ for constant $\rho$ surface. 
    We can see $\Psi^{(2)}$ rapidly goes to zero at some temperature. 
    This shows the condensation of the scalar field under a critical temperature, which corresponds to the condensation of the Cooper pair during the superconducting phase in QFT.

\section{Linear perturbation on the holographic superconductor}
\label{sec:perturbation}

    We will consider the charged scalar and the $U(1)$ gauge field perturbations on the background. The background solutions are obtained in the previous section. We impose the boundary conditions for perturbation fields as if there exists a point source for the $U(1)$ gauge field on the AdS boundary at the south pole. Then we observe its response through a lens at some observation point on the AdS boundary, as we depicted in Fig.~\ref{fig:AdS_boundary}. 

    The linear perturbative equations of motion are
    \begin{equation}
        \label{eq:perturbation_EoM_scalar}
        \mathcal{D}^2\delta \Psi +\frac{2}{L^2}\delta \Psi=2i \delta A^\mu\mathcal{D}_\mu\Psi+i(\nabla_\mu\delta A^\mu)\Psi
    \end{equation}
    \begin{equation}
        \label{eq:perturbation_EoM_gauge}
        \nabla_\nu \delta F^{\nu\mu}=\delta J^\mu,
        \ \ \ 
        \delta J^\mu = i(\delta \Psi^* \mathcal{D}^\mu \Psi - \delta \Psi (\mathcal{D}^\mu \Psi)^*  + \Psi^* \mathcal{D}^\mu \delta \Psi  - \Psi (\mathcal{D}^\mu \delta\Psi)^* ) + 2\Psi^2 \delta A^\mu.
    \end{equation}
    $\Psi$ and $\Phi$ are the background solution as obtained in the previous section and $\mathcal{D}_\mu=\partial_\mu+i \Phi \delta^t_{\ \mu}$ is the gauge covariant derivative. 
    We decompose the perturbation into the vector and scalar modes.
    Since we will only consider the axisymmetric external source in the boundary theory, 
    we focus on axisymmetric linear perturbation: $\partial_\varphi \delta A_\mu=\partial_\varphi \delta\Psi = 0$.

\subsection{Vector mode perturbation}

    In the following, we will study the vector mode perturbation. 
    In $S^2$, the vector spherical harmonics $(Y_{lm}(\theta))_i$  simply relate to the scalar spherical harmonics $Y_{lm}(\theta)$ as
    \begin{equation}
        (Y_{lm}(\theta,\varphi))_i = \epsilon_{ij} \hat{D}^j Y_{lm}(\theta,\varphi)\quad(l\geq 1)\ ,
    \end{equation}
    where $\hat{D}$ and $\epsilon_{ij}$ are the covariant derivative and the Levi-Civita tensor on the unit $S^2$.
    We can decompose the axisymmetric vector mode by the vector spherical harmonics as
    \begin{equation}
        \label{eq:vector_mode_field}
        \delta A_a=0\ ,\quad
        \delta A_i = e^{-i \omega t} \sum_{l=1}^\infty c_l \alpha^{l}(r) (Y_{l0}(\theta))_i\ ,\quad
        \delta\Psi = 0\ ,
    \end{equation}
    where $a,b=t,r$ and $i,j=\theta,\varphi$. Here, we consider the monochromatic wave with the frequency $\omega$. 
    The constant $c_l$ is defined in Eq.~(\ref{eq:vector_mode_cl}). 
    The equation of motion for $\alpha^l(r)$ is given by the Schr\"{o}dinger form as
    \begin{align}
        \label{eq:vector_mode_eom}
        \left[ -\frac{d^2}{dr_*^2} + \left( \frac{l(l+1)}{r^2} + 2 \Psi^2 \right) F(r_*) \right] \alpha^{l}(r_*) = \omega^2 \alpha^{l}(r_*)\ .
    \end{align}
    We impose the in-going wave condition at the horizon $d\alpha^{l}/dr_*|_{r=r_h}=-i\omega \alpha^{l}|_{r=r_h}$.
    We also set the normalization condition at the AdS boundary $\alpha^l|_{r=\infty}=1$.
    Then, the asymptotic form of the gauge field $\delta A_i$ becomes
    \begin{equation}
        \delta A_i|_{r=\infty} =e^{-i \omega t} \sum_{l=1}^\infty c_l (Y_{l0}(\theta))_i = e^{-i \omega t}  \epsilon_{ij}\hat{D}^j (\sum_{l=1}^\infty c_l Y_{l0}) =  e^{-i \omega t}  \epsilon_{ij}\hat{D}^j g(\theta)\ ,
        \label{Avecinf}
    \end{equation}
    where $g(\theta)$ is the Gaussian function defined in Eq.~(\ref{eq:vector_mode_cl}). 
    According to the AdS/CFT dictionary, $\delta A_i|_{r=\infty}$ correspond to the external source in the boundary theory.
    The above equation implies that the spatially localized time-periodic external electromagnetic field is applied to the superconductor on $S^2$.

    The asymptotic solution of $\alpha^l(r)$ near the AdS boundary is 
    \begin{equation}
        \label{eq:vector_mode_asmp}
        \alpha^l(r\to\infty)
        =1+\frac{\alpha^{l(1)}}{r}+\cdots\ .
    \end{equation}
    Then the response (i.e., electric current) for the vector mode is given as follows:
    \begin{equation}
        \label{eq:vector_mode_response}
        \langle J^V_i(t,\theta)\rangle
        = e^{-i \omega t} \sum_{l=1}^\infty c_{l} \alpha^{l(1)} \left(Y_{lm}(\theta)\right)_i
        = e^{-i \omega t} \epsilon_{ij}\hat{D}^j  \langle J^V(\theta)\rangle\ ,
    \end{equation}
    where
    \begin{equation}
        \langle J^V(\theta)\rangle \equiv  \sum_{l=1}^\infty c_{l} \alpha^{l(1)} Y_{lm}(\theta) + \textrm{const}\ .
    \end{equation}
    See Appendix \ref{appendix_response} for details. 
    The constant term in $\langle J^V(\theta)\rangle$ is ambiguous.
    We determines it so that $\int_0^d d\theta \sin\theta \langle J^V(\theta)\rangle = 0$ is satisfied.
    We will use $\langle J^V(\theta) \rangle$  to construct the image of the black hole.
    We will numerically solve Eq.~\eqref{eq:vector_mode_eom} to obtain the response.

\subsection{Scalar mode perturbation}

    We will focus on the scalar mode perturbative solutions. We decompose the scalar mode by the scalar harmonics as 
    \begin{align}
        \label{eq:scalar_mode_gauge_field}
        &\delta A_a
        =\sum_{l=1}^\infty c_l a_a^{l}(t,r) Y_{l0}(\theta),\\
        &\delta A_i = \sum_{l=1}^\infty c_l \beta^{l}(t,r) \hat{D}_i Y_{l0}(\theta),\label{eq:scalar_mode_beta}\\
        &\delta\Psi
        =\sum_{l=1}^\infty c_l \psi^{l}(t,r) Y_{l0}(\theta)\ ,
        \label{eq:scalar_mode_scalar_field}
    \end{align}
    where $c_l$ is the constant defined in Eq.~(\ref{eq:vector_mode_cl}). Since the Maxwell field does not have dynamical degrees of freedom for $l=0$, we only consider $l\geq 1$. We will impose the gauge condition as $\beta^{l}(t,r)=0$. 
    Notice that since we consider $2$ dimensional vector space, the $l$-component of the field strength $f^l_{ab}(t,r)$ has only one degree of freedom. That is, we can describe it by the scalar function $f^l(r)$ and the complete asymmetric tensor density $\epsilon_{ab}$ as
    \begin{equation}
    \label{eq:elemag_tensor_scalar}
        f^l_{ab}(t,r)
        = \partial_a a^l_b(t,r)-\partial_b a^l_a(t,r)
        =\frac{1}{r^2} f^l(t,r)\epsilon_{ab}\ .
    \end{equation}
    Then, in the following analysis, we will deal with $f^l(t,r)$ instead of $a_a^{l}(t,r)$ itself. The equation of motions for $f^l(t,r)$, $\psi^l(t,r)$ and $\psi^{*l}(t,r)$ are summarized as
    \begin{align}
        & D_a\left[C(r)\left(D^a f^l(t,r)+\epsilon^{ab}r^2 j^l_b(t,r) \right)\right]=\frac{f^l(t,r)}{r^2},\label{eq:scalar_mode_Maxwell_b} \\
        &\left\{
        \mathcal{D}_a\mathcal{D}^a + \frac{2}{r}g^{ar}\mathcal{D}_a - \frac{l(l+1)}{r^2} + \frac{2}{L^2}
        \right\}\psi^l(t,r) \nonumber\\
        &\qquad
        =\left\{
        2iC(r)\left(\epsilon^{ba}D_b f^l-r^2 (j^l)^a\right) \left(\mathcal{D}_a+\frac{\partial_a r}{r}\right)
        +\Psi^*\psi^l-\Psi\psi^{*l}\right\}\Psi\ .
        \label{eq:scalar_mode_KG}
    \end{align}
    Here, $D_a$ is the covariant derivative with respect to the $(t,r)$-part of the metric and  
    \begin{equation}
        \label{eq:scalar_mode_j_a}
        j^l_a(t,r) \equiv i\left(\psi^{*l}\mathcal{D}_a\Psi-\psi^l (\mathcal{D}_a\Psi)^*+\Psi^*\mathcal{D}_a\psi^l-\Psi(\mathcal{D}_a\psi^l)^*\right)\ ,
    \end{equation}
    \begin{equation}
        \label{eq:scalar_mode_C(r)}
        C(r)\equiv \frac{1}{\lambda_l+2 r^2 |\Psi|^2}\ .
    \end{equation}
    Details of derivation are in Appendix \ref{appendix_scalarEOM}. We now assume the time dependence of the perturbation variable as $f^l(t,r)=e^{-i\omega t}f^l(r)$, 
    $\psi^l(t,r)=e^{-i\omega t}\psi^l(r)$ and  $\psi^{\ast}{}^l(t,r)=e^{-i\omega t}\psi^\ast{}^l(r)$.
    We impose the in-going wave condition at the horizon $r=r_h$ as
    \begin{equation}
        \frac{\partial f^l}{\partial r_*}=-i \omega  f^l\ ,\quad 
        \frac{\partial \psi^l}{\partial r_*}=-i \omega  \psi^l\ ,\quad 
        \frac{\partial \psi^\ast{}^l}{\partial r_*}=-i \omega  \psi^\ast{}^l\ .
    \end{equation}
    At infinity, the asymptotic behaviors of $f^l(r)$, $\psi^l(r)$  and $\psi^\ast{}^l(r)$ as
    \begin{equation}
        \begin{split}
        &f^l=f^{l(0)}+\frac{f^{l(1)}}{r}+\frac{f^{l(2)}}{r^2}+\cdots\ ,\\
        &\psi^l=\frac{\psi^{l(1)}}{r}+\frac{\psi^{l(2)}}{r^2}+\cdots\ ,\quad 
        \psi^\ast{}^l=\frac{\psi^\ast{}^{l(1)}}{r}+\frac{\psi^\ast{}^{l(2)}}{r^2}+\cdots\ .
        \end{split}
        \label{eq:scalar_mode_asmp}
    \end{equation}
    As shown in Appendix~\ref{appendix_response}, the leading term $f^{l(0)}$ is proportional to the response and the second leading term $f^{l(1)}$ corresponds to the external electromagnetic field. 
    This is contrary to the vector mode.
    Then, we will solve Eqs.~\eqref{eq:scalar_mode_Maxwell_b} and \eqref{eq:scalar_mode_KG} on the following boundary condition.
    \begin{equation}
        \label{eq:scalar_mode_bc4}
        f^{l(1)}=-l(l+1),\ \ \ 
        \psi^{l(1)}=\psi^\ast{}^{l(1)}=0
    \end{equation}
    It is also shown in Appendix \ref{appendix_scalarEOM} that, near the infinity,  we can reproduce $a^l_a(t,r)$ from $f^l(t,r)$ as 
    \begin{equation}
        \label{eq:scalar_mode_a(f)}
        a^l_a(t,r)\simeq \frac{1}{l(l+1)}\epsilon_{ba}D^b f^l(t,r)\ .
    \end{equation}
    The first condition in Eq.~(\ref{eq:scalar_mode_bc4}) is equivalent to the boundary conditions $a^l_t|_{r=\infty} = 1$. 
    Therefore, as the external electromagnetic field in the superconductor, we have
    \begin{equation}
        \label{eq:scalar_mode_source}
        \delta A_t|_{r=\infty}
        =e^{-i \omega t}\sum_{l=1}^\infty c_l Y_{l0}(\theta)
        =e^{-i \omega t}(g(\theta)-g_0)\ ,
    \end{equation}
    where $g_0=\int_0^\pi \sin\theta \,g(\theta)/2$.
    The responses are obtained by
    \begin{align}
        &\langle J^S_t(t,\theta) \rangle  
        =- e^{-i \omega t} \sum_{l} c_{l} f^{l(0)} Y_{l0}(\theta)
        = e^{-i \omega t}  \langle J^S(\theta) \rangle\ ,\label{eq:scalar_mode_response_t} \\
        &\langle J^S_i(t,\theta) \rangle  
        =i\omega e^{-i \omega t} \sum_{l} \frac{c_l}{l(l+1)} f^{l(0)} \hat{D}_i Y_{l0}(\theta)\ ,\label{eq:scalar_mode_response_i}
    \end{align}
    where 
    \begin{equation}
        \langle J^S(\theta) \rangle  \equiv -\sum_{l=0}^\infty c_{l} f^{l(0)} Y_{l0}(\theta)\ .
    \end{equation}
    See Appendix \ref{appendix_response} for the detail.
    They are related by the charge conservation: 
    \begin{equation}
        \partial_t \langle J^S_t(t,\theta) \rangle + \hat{D}^i\langle J^S_i(t,\theta) \rangle = 0\ .
    \end{equation}
    We will use $\langle J^S(\theta) \rangle$ for constructing the image of the black hole.

\subsection{Imaging black hole from the dissipation part of the response function}

    Response functions are complex valued functions. Both of their real and imaginary parts are observable.
    For the vector mode, the external electric field in $S^2$ is given by
    \begin{equation}
    E_i = -\delta F_{ti}= -\partial_t \delta A_i |_{r=\infty}=  \epsilon_{ij}\hat{D}^j g(\theta) \sin \omega t\ ,
    \end{equation}
    where we took the real part of Eq.~(\ref{Avecinf}). 
    As its response, we obtain the electric current as
    \begin{equation}
        \langle J^V_i(t,\theta)\rangle =  \epsilon_{ij}\hat{D}^j \left[\textrm{Re} \langle J^V(\theta)\rangle \cos\omega t +  \textrm{Im} \langle J^V(\theta)\rangle \sin\omega t \right]\ .
    \end{equation}
    Again we used the real part of Eq.~(\ref{eq:vector_mode_response}). 
    The first term in the electric current has the phase difference $\pi/2$ with respect to the electric current. (Borrowing the terminology from electric circuit theory, 
    we can regard $\textrm{Re} \langle J^V(\theta)\rangle$ and $\textrm{Im} \langle J^V(\theta)\rangle$ as the ``reactance'' and  ``resistance'', respectively.)
    Only $\textrm{Im} \langle J^V(\theta)\rangle$ contributes to the Joule heating, $Q=\lim_{T\to \infty}\int_{-T}^{T} dt E^i \langle J^V_i \rangle/(2T)$. 
    The same applies to the scalar mode. We will refer $\textrm{Im}\langle J^{V,S}(\theta) \rangle$ as dissipation parts of the response functions since the ``real'' and ``imaginary'' parts do not have proper meanings. (We can exchange their roles by the constant shift of the time, $t\to t+\pi/(2\omega)$.)

    In the previous works~\cite{Hashimoto:2018okj,Hashimoto:2019jmw}, the response function is directly used for constructing the image of the black hole as explained in section.\ref{sec:Review}.
    In this paper, we propose a new prescription for constructing a clear image of the black hole: We use
    dissipation parts of the response functions for constructing the image of the black hole, i.e. $\mathcal{M} (\vec{x})=\textrm{Im}\langle J^{V,S}(\theta) \rangle$ in Eq.~(\ref{eq:lens_formula}).
    We can understand that the dissipation part of the response has  clear information about the photon sphere as follows.
    Let us consider the response function in $l$-space, $\alpha^{l(1)}$ and $f^{l(0)}$.
    In the normal phase, $\Psi(r)=0$, perturbation equations for vector and scalar modes are identical and given by Schr\"{o}dinger form~(\ref{eq:vector_mode_eom}).
    When the ``energy'' $\omega^2$ is larger than the top of the potential $l(l+1)v_\textrm{max}$, 
    the wave sent from the AdS boundary directly plunges into the black hole and there should be a non-negligible dissipation (or Joule heating in the QFT).
    Thus, Im$\alpha^{l(1)}$ and Im$f^{l(0)}$ are also non-negligible. On the other hand, when $\omega^2$ is smaller than the top of the potential, 
    Im$\alpha^{l(1)}$ and Im$f^{l(0)}$ are suppressed by the tiny tunneling probability. 
    It follows that they suddenly become small at $l \simeq \omega/\sqrt{v_\textrm{max}}$. This coincides with the angular momentum of the null geodesic on the photon sphere. (See Eq.~(\ref{lphoton}).)
    In Appendix \ref{appendix_WKB}, we did the detailed WKB analysis and 
    showed that the photon ring appears at (\ref{Estnradius}) in the image of the black hole constructed by the formula~(\ref{eq:lens_formula}).

\section{Imaging holographic superconductor}
\label{sec:imaging_sc}

     We will show our results of the image of the holographic superconductor system.
    Let us observe $\mathcal{M}$ at some observation point $\theta=\theta_{\rm{obs}}$ on the AdS boundary as we depicted in Fig.~\ref{fig:AdS_boundary}. We will apply the Fourier-Fresnel transformation in Eq.~\eqref{eq:lens_formula} to the observables to construct the image.

\subsection{Image of vector-mode perturbation}

\begin{figure}
    \centering
    \includegraphics[width=150mm]{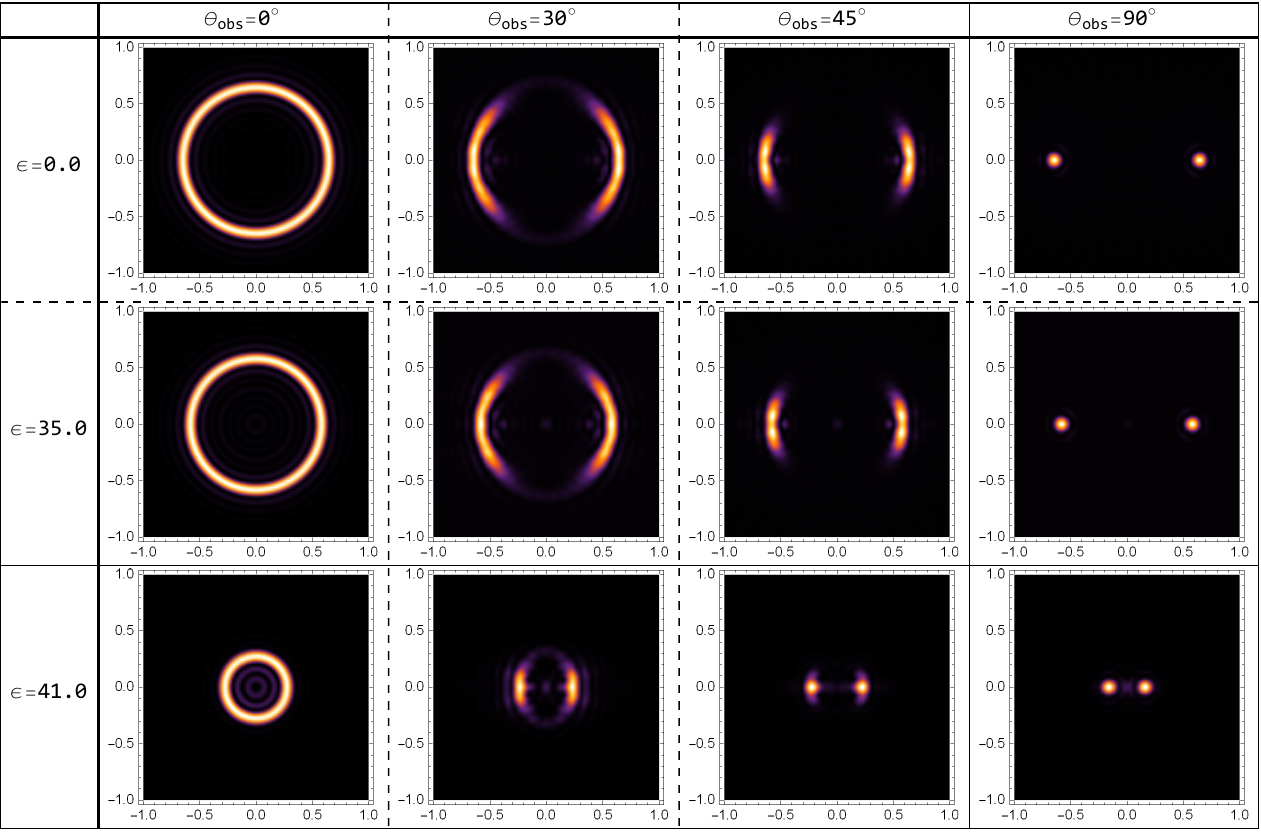}
    \caption{The image for a vector-mode perturbation. We calculate the image of a dissipation part of the response, where $r_h=0.3$, $\omega=80$, $d=0.5$, and $\sigma=0.01$.}
    \label{fig:BHimage_vector_Im}
\end{figure}

    We depicted the image of the vector-mode gauge field perturbation in Fig.~\ref{fig:BHimage_vector_Im}. 
    The horizontal line and the vertical line are $x_s/f$ and $y_s/f$ respectively. We set the horizon radius $r_h=0.3$, frequency $\omega=80$, a lens radius $d=0.5$, and  variance of Gaussian source $\sigma=0.01$. 
    A background scalar value at the horizon varies as $\epsilon=0.0, 35.0, 41.0$, and the observation point varies as $\theta_{\rm{obs}}=0^\circ, 45^\circ, 60^\circ,90^\circ$. 
    In terms of the boundary values, we will build the images corresponding to $\mu(\epsilon)=2.533, 130.5, 153.1$ or $\rho(\epsilon)=0.7537, 5132, 7068$, which can be read from Fig.~\ref{fig:mu_epsilon} and \ref{fig:rho_epsilon} respectively.

\begin{figure}
    \centering
    \includegraphics[width=80mm]{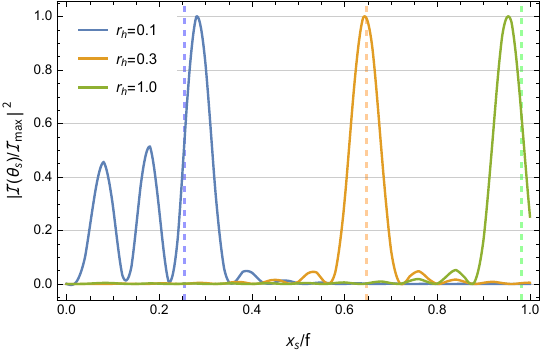}
        \caption{The $r_h$ dependence of a cross-section view of the image for a vector-mode perturbation. Each dashed line represents a photon sphere radius calculated from the geodesic approximation.}
        \label{fig:BHimage_vector_rh}
\end{figure}

	We obtain axisymmetric 
 images when $\theta_{\rm{obs}}=0^\circ$, which is trivial from the axisymmetry of the system. However, each ring radius is non-trivially determined from the details of the dual bulk. 
	According to \eqref{Estnradius} and \eqref{eq:vector_mode_eom}, we can estimate photon ring radius from an effective potential given by
    \begin{equation}
        \label{eq:effective_potential}
        V_{\rm{eff}}(r_*;l,\epsilon)=\left(\frac{l(l+1)}{r^2}+2\Psi^2\right)F(r_*)\ .
    \end{equation}
As shown in Appendix.\ref{ImagingfromIm}, in the WKB approximation, 
the photon ring radius is determined by $x_S/f = l_0/\omega$ where $l=l_0$ is chosen so that the muximum value of the effective potential equals to $\omega^2$,  i.e, $\omega^2={\rm{max}}\,V_{\rm{eff}}(r_\ast;l_0,\epsilon)$.
In Fig.~\ref{fig:BHimage_vector_rh}, we depicted a cross-section view of the image
with solid lines and the photon ring radius calculated from the WKB analysis with dashed lines with respect to $r_h=0.1,0.3,1.0$ with $\omega=80, \epsilon=10.0, d=0.5$ and $\sigma=0.01$. The vertical line is the image amplitude which is normalized by the maximum amplitude, and the horizontal line is $x_s/f$. It is a noteworthy fact that the ring radius of images almost coincide with the photon ring radius estimated by the WKB approximation.
    
    When constructing the image applying \eqref{eq:lens_formula} to the weakly coupled theories, 
for example, in the case of $\phi^4$-theory, the ring radius is given by $\theta_{obs}=\pi/2$ regardless of the temperature when $\omega$ is sufficiently large~\cite{Hashimoto:2019jmw}. Then, we can qualitatively distinguish whether QFT has its gravity dual or not by analyzing the temperature dependence of the image. Hence, coincidence with the photon ring and temperature dependence of the ring images in Fig.~\ref{fig:BHimage_vector_Im} represent not only an axisymmetric property of the system but also properties of their gravity dual.

    During the normal conducting phase $\epsilon=0.0$, we see a ring image with its radius $r_s\sim 0.646$ at $\theta_{\rm{obs}}=0^\circ$. 
    According to  Eq.~\eqref{Estnradius}, the photon ring radius is given by $r_s \simeq 0.647$. Therefore, we conclude that the higher-order Einstein ring winding around the vicinity of the photon sphere emerges in the image during the normal conducting phase.
    As the observation point gets closer to the equator $\theta_{\rm{obs}}=90^\circ$, the image transforms from the ring image to 2 bright points. These 2 bright points represent the wave propagating clockwise and counterclockwise on the $\varphi=0^\circ$ and $180^\circ$ surfaces.

    During the superconducting phase $\epsilon >0.0$, there are some interesting transitions in the image. First, the ring radius shrinks little by little for $0 < \epsilon \lesssim 35$. This represents the 2nd order phase transition of the superconductor. For $35 \lesssim \epsilon$, the ring radius shrinks drastically. 

\begin{figure}
    \centering
    \includegraphics[width=150mm]{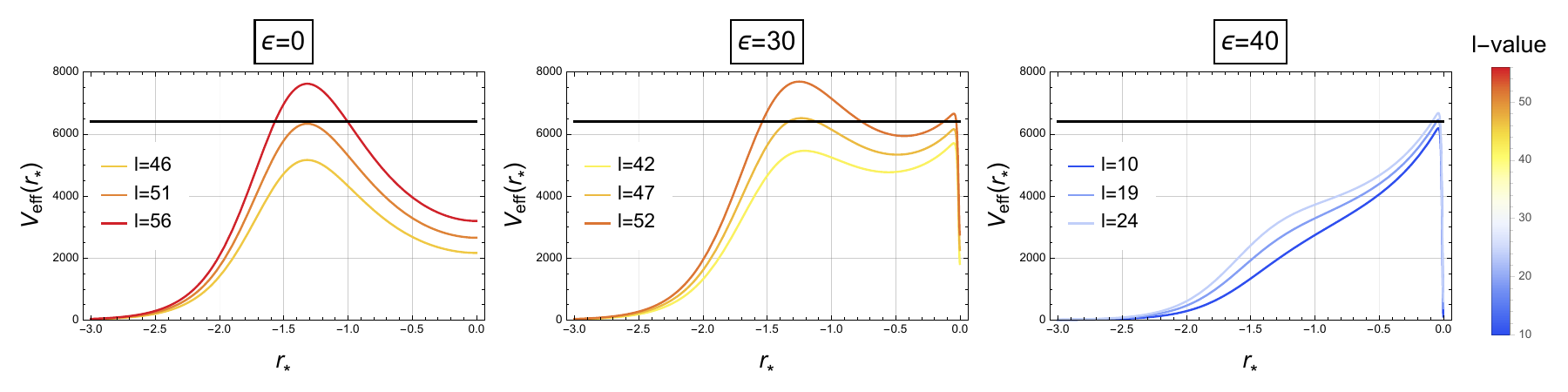}
    \caption{The behaviour of the effective potential with respect to tortoise coordinate: $V_{\rm{eff}}(r_*;l\epsilon)$. We set $r_h=0.3, \omega=80$, and $\epsilon=0.0, 30.0, 40.0$. A black solid line represents $\omega^2$.}
    \label{fig:effective_potential}
\end{figure}

    The change of images around $\epsilon \sim 35$ can be understood by considering the effective potential $V_{\rm{eff}}$ in \eqref{eq:effective_potential}. (See Fig.~\ref{fig:effective_potential}.) 
    The first term on the right hand side in \eqref{eq:effective_potential} is the gravitational potential, and the second term represents the contribution from the background scalar to the potential. Each term corresponds to the left side potential hill and the right side potential hill, respectively, in Fig.~\ref{fig:effective_potential}. 
    When $\omega^2>V_{\rm{eff}}(r_*;l,\epsilon)$ is satisfied for any $r_*$, an electromagnetic field falls into the black hole, so the ring images of these $l$-components do not emerge in the image.  
    When $\omega^2 \leq V_{\rm{eff}}(r_*;l,\epsilon)$ is satisfied at the maximum of the potential, null rays emitted from the AdS boundary do not fall into the black hole. 
    Particularly when $\omega^2 = V_{\rm{eff}}(r_*;l,\epsilon)$, null rays propagate in a circular motion at the top of the potential, which we call an effective photon sphere. As discussed in \cite{Hashimoto:2018okj,Hashimoto:2019jmw} the contribution of the null rays winding around the vicinity of the effective photon sphere is much larger than that with a smaller winding number. Hence, we can roughly expect to see the ring image corresponding to $l$ mode, which satisfies $\omega^2 = V_{\rm{eff}}(r_*;l,\epsilon)$ at the top of the potential.

    During $0 < \epsilon \lesssim 35$, the gravitational potential is higher than the background scalar potential. The gravitational potential is lifted  by the background scalar field as we raise $\epsilon$. Then, $\omega^2$ grazes the top of the effective potential even for small $l$. As we saw in Section~\ref{sec:Review}, a small $l$ wave emerges as a small ring. Therefore, the photon ring shrinks gradually as we raise the amplitude of a background scalar field in $0 < \epsilon \lesssim 35$.

    On the other hand, the background scalar potential is higher than the gravitational potential when $35 \lesssim \epsilon$. Therefore, a drastic change in the image around $\epsilon \sim 35$ is originated from such a transition of the potential top  Although the ring radius is expected to shrink gradually after the transition as we raise $\epsilon$, we could not verify such behavior since the image amplitude falls below the numerical error.

\begin{figure}
    \centering
    \includegraphics[width=90mm]{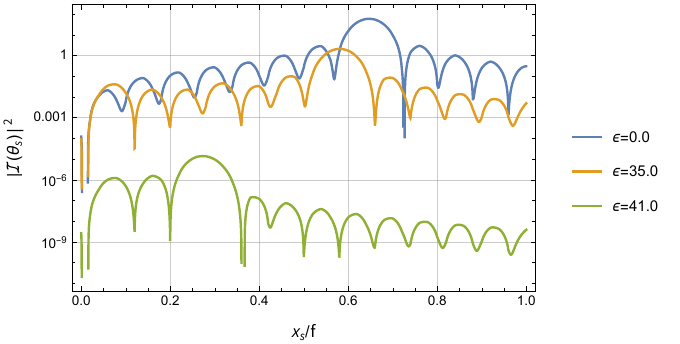}
    \caption{The $\epsilon$ dependence of a cross-section view of the image for a vector-mode perturbation. We set $r_h=0.3$, $\theta_{\rm{obs}}=0^\circ$, $\omega=80$, $d=0.5$, $\sigma=0.01$, and $\epsilon=0.0, 35.0, 41.0$.}
    \label{fig:BHimage_vector_epsilon}
\end{figure}

    We showed a cross-section view of the black hole image for $\theta_{\rm{obs}}=0$ in Fig.~\ref{fig:BHimage_vector_epsilon}. The vertical axis is the amplitude of the image, and the horizontal axis is $x_s/f$. We set $r_h=0.3$, $\theta_{\rm{obs}}=0^\circ$, $\omega=80$, $d=0.5$  $\sigma=0.01$, and $\epsilon$ varies as $\epsilon=0.0, 35.0, 41.0$. The amplitude decays as we raise $\epsilon$ since the background potential barrier grows and obstruct a wave propagated from the AdS boundary. 

\begin{figure}
    \centering
    \includegraphics[width=80mm]{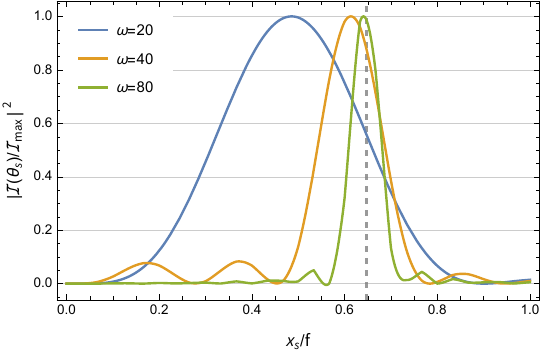}
     \caption{The $\omega$ dependence of a cross-section view of the image for a vector-mode perturbation. A gray dashed line represents a photon sphere radius calculated from the geodesic approximation.}
        \label{fig:BHimage_vector_omega}
\end{figure}

    We also depicted the amplitude of the black hole image with respect to the change in a wave frequency in Fig.~\ref{fig:BHimage_vector_omega}. We set $r_h=0.3$, $\omega=20,40,80$, $\epsilon=10.0$, $\theta_{\rm{obs}}=0$, $d=0.5$ and $\sigma=0.01$.
    The vertical line is the image amplitude which is normalized by the maximum amplitude, and the horizontal line is $x_s/f$. A gray dashed line represents the photon sphere radius calculated from the null geodesic approximation. The ring image becomes blurred as we decrease $\omega$ due to the wave effect. We can see $x_s/f$ of the maximum amplitude gets closer to the photon sphere radius calculated from a null ray approximation as we raise $\omega$.

\subsection{Scalar-mode perturbation}

\begin{figure}
    \centering
    \includegraphics[width=150mm]{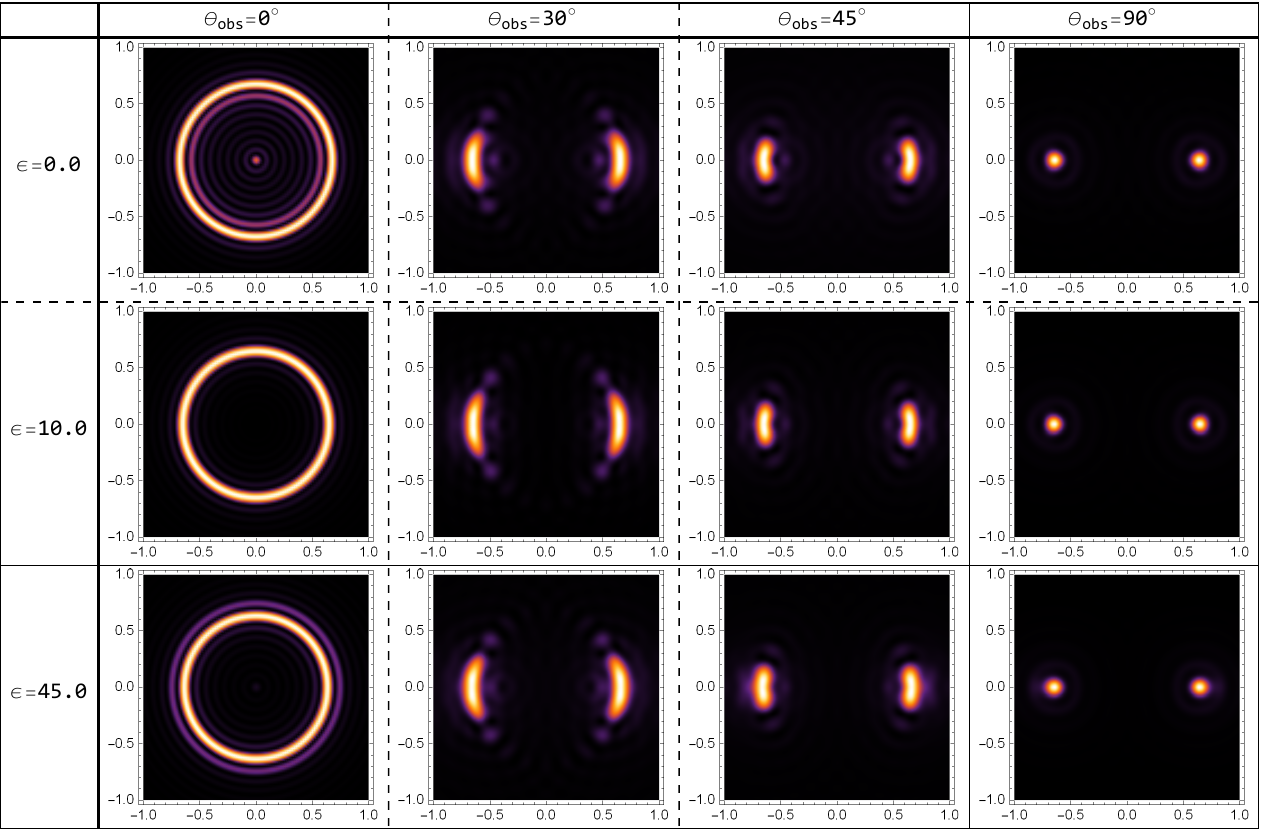}
    \caption{The image for a scalar-mode perturbation. We calculate the image of a dissipation part of the response, where $r_h=0.3$, $\omega=80$, $d=0.5$, and $\sigma=0.01$.}
    \label{fig:BHimage_scalar_Im}
\end{figure}

    We showed the image of a scalar-mode gauge field perturbation in Fig.~\ref{fig:BHimage_scalar_Im}. The horizontal line and the vertical line are $x_s/f$ and $y_s/f$ respectively. We set the horizon radius $r_h=0.3$, frequency $\omega=80$, the lens radius $d=0.5$, and the variance of the Gaussian source $\sigma=0.01$. A background scalar value at the horizon $\epsilon$ varies as $\epsilon=0.0,10.0,45.0$, and the observation point varies as $\theta_{\rm{obs}}=0^\circ,45^\circ,60^\circ,90^\circ$. 
    In terms of the boundary values, we will build the images corresponding to $\mu(\epsilon)=2.533, 36.26, 168.1$ or $\rho(\epsilon)=0.7537, 391.5 ,8530$, which can be read from Fig.~\ref{fig:mu_epsilon} and \ref{fig:rho_epsilon} respectively.

    As for a vector-mode perturbation, we see a photon ring at $\theta_{\rm{obs}}=0^\circ$ during the normal phase $\epsilon=0.0$. Also, as we vary the observation point from $0^\circ$ to $90^\circ$, each ring image tends to be 2 bright points, which is also the same as a vector-mode perturbation qualitatively.
    
\begin{figure}[ht]
	\begin{tabular}{cc}
	\begin{minipage}{.45\textwidth}
        \centering
       \includegraphics[width=65mm]{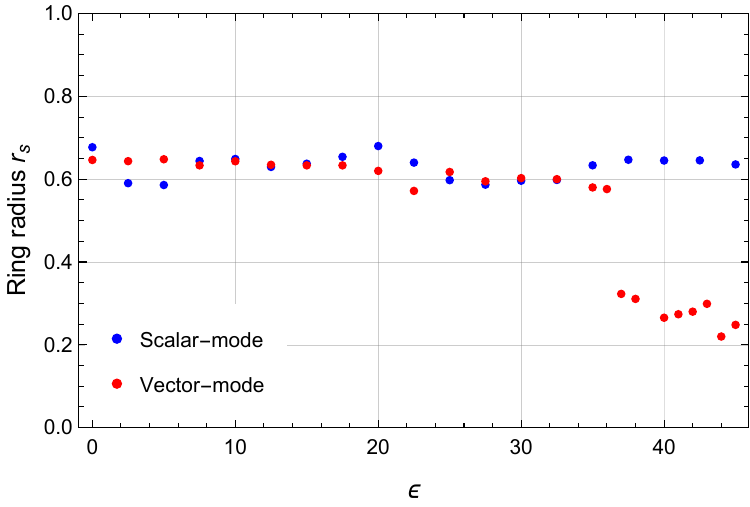}
        \caption{The $\epsilon$ dependence of a ring radius in the image. Blue points and red points correspond to a scalar-mode and a vector-mode perturbation respectively.}
        \label{fig:BHimage_radius}
	\end{minipage}
	\hspace{3mm}
	\begin{minipage}{.45\textwidth}
        \centering
        \includegraphics[width=65mm]{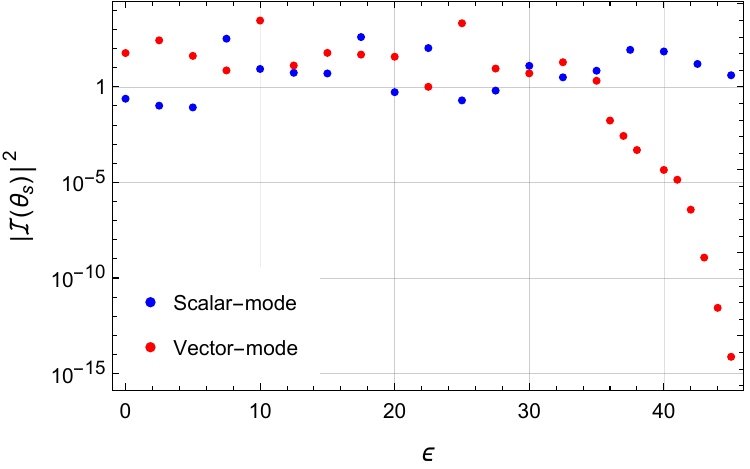}
        \caption{The $\epsilon$ dependence of an image amplitude. Blue points and red points correspond to a scalar-mode and a vector-mode perturbation respectively.}
        \label{fig:BHimage_amplitude}
	\end{minipage}
	\end{tabular}
\end{figure}

    There are 2 critical differences in the image between a scalar-mode and a vector-mode perturbation, which are both originated by the coupling of a gauge field perturbation and scalar field perturbation in \eqref{eq:scalar_mode_Maxwell_b} and \eqref{eq:scalar_mode_KG}.
    First, the ring image does not shrink during the superconducting phase. We showed a ring radius with respect to $\epsilon$ for a scalar-mode and a vector-mode perturbation in Fig.~\ref{fig:BHimage_radius}. We set $r_h=0.3$, $\theta_{\rm{obs}}=0^\circ$, $\omega=80$, $d=0.5$, and $\sigma=0.01$. For a vector-mode perturbation, we see a radius gets smaller as we raise $\epsilon$, and it shrinks dramatically around $\epsilon \sim 35$. In contrast, a radius does not alter for a scalar-mode perturbation. We expect a gauge field is scattered by a scalar field perturbation, and observe the effective photon ring with a larger radius compared to a vector-mode perturbation. 
    Second, the image amplitude does not decay as we raise $\epsilon$. We depicted the $\epsilon$ dependence of  an image amplitude in Fig.~\ref{fig:BHimage_amplitude} for a scalar-mode and a vector-mode perturbation. We set $r_h=0.3$, $\theta_{\rm{obs}}=0^\circ$, $\omega=80$, $d=0.5$, and $\sigma=0.01$. For a vector-mode perturbation, the amplitude decays as $\epsilon$ becomes larger since the background potential barrier screens the wave propagated from the AdS boundary. For a scalar-mode perturbation, we guess that a gauge field obtains the energy to exceed such a potential barrier due to its excitation by scalar-field. It is difficult to understand these behaviors quantitatively, for example by considering an effective potential, since the equations of motion for scalar-mode perturbations are complicatedly coupled.

\section{Conclusion}
\label{sec:Conclusion}

    We proposed a way to take the image of the black hole that is dual to a superconductor.
    We considered an external time-periodic localized electromagnetic field in the superconductor and its response (i.e., electric current).
    We applied the Fourier-Fresnel transformation~(\ref{eq:lens_formula})  to the response function after multiplying the window function. 
    Then, we only considered the dissipation part (or the imaginary part in our convention) of the response function to take the clear image of the black hole.
    Typical images were summarized in Figs.~\ref{fig:BHimage_vector_Im} and \ref{fig:BHimage_scalar_Im}.
    We also estimated the radius of the photon ring in the image as a function of the scalar condensate $\epsilon\equiv \Psi(r_h)$.
    For the vector mode, we found the discontinuous change of the radius for a sufficiently large $\epsilon$.
    On the other hand, for the scalar mode, we did not find its discontinuous change. Then, the radius does not depend on $\epsilon$ much as far as we studied. 
    For the scalar mode, on the other hand, the radius  does not depend on $\epsilon$ so much and we did not find its discontinuous change as far as we studied.
    Our results indicate that we can observe black holes by the tabletop experiment of superconductors if they have gravitational duals.
    The observation of black holes can be used as the experimental test of the existence of the gravitational dual for given materials.

    We only considered the probe limit of the holographic superconductor. 
    The holographic model with the back reaction to the metric has been considered in Ref.\cite{Hartnoll:2008kx}.
    We can also apply the imaging of the black hole to such a model. 
    It is also interesting to consider the application to the p- or d-wave holographic superconductor models~\cite{Gubser:2008wv,Chen:2010mk,Benini:2010pr}.

    The other future direction is to apply our method to the Sachdev–Ye–Kitaev (SYK) model~\cite{Sachdev:1992fk,Kitaev-talk}.
    Originally, the SYK model was introduced as the (0+1)-dimensional model but extended to (1+1)-dimensional spacetime~\cite{Turiaci:2017zwd,Jian:2017unn,Das:2017pif,Murugan:2017eto,Das:2017hrt}.
    The (1+1)-dimensional model is probably a dual of the (2+1)-dimensional black hole.
    It is interesting to consider the imaging of the black hole dual to the SYK model both on the gravity and field theory sides. 
    There is also an attempt to realize the SYK model in a real experiment~\cite{Danshita:2016xbo}. 
    It would be nice if we can observe the black hole through such an experiment.

\acknowledgments
    We would like to thank Takaaki Ishii and Chul-Moon Yoo for useful conversations.
    The work of K.~M.~was supported in part by JSPS KAKENHI Grant Number JP18H01214 and JP20K03976.

\appendix
\section{Derivation of equations of motion for a scalar-mode perturbation}
\label{appendix_scalarEOM}
    We will explain the derivation of \eqref{eq:scalar_mode_Maxwell_b} and \eqref{eq:scalar_mode_KG} in this section.
    We start from perturbative Maxwell equations \eqref{eq:perturbation_EoM_gauge} and the equations of motion for charged scalar field \eqref{eq:perturbation_EoM_scalar} which $a_a(t,r)$, $\psi(t,r)$ and $\psi^*(t,r)$ follow.

    First, we focus on the $\mu=b$ component of the Maxwell equation:
    \begin{equation}
    \label{eq:Maxwell_b}
        \nabla_\nu \delta F^{\nu b}=\delta J^b
    \end{equation}
    The right hand side is defined as
    \begin{equation}
        \delta J^a =\sum_l c_l \left(j^a(t,r) +2|\Psi|^2 a^a(t,r)\right) Y_{l0}(\theta)\ ,
    \end{equation}
    where $j^a(t,r)$ is given in \eqref{eq:scalar_mode_j_a}.
    The left hand side will be 
    \begin{equation}
        \nabla_\nu \delta F^{\nu b}=
        \frac{1}{r^2}h^{bc} \sum_l c_l \left\{
        D^a\left(r^2 f^l_{ac}(t,r)\right)-l(l+1) a^l_c(t,r)
        \right\}Y_{l0}(\theta)\ ,
    \end{equation}
    where we defined $f^l_{ab}(t,r)$ in \eqref{eq:elemag_tensor_scalar}. 
    By comparing the coefficient of $Y_{l0}$ on the both hand sides, we get
    \begin{equation}
         D^a\left(r^2f^l_{ab}(t,r)\right)-\left\{l(l+1)+2r^2|\Psi|^2\right\} a^l_b(t,r) = r^2 j_b\ .
        \label{scalareq}
    \end{equation}
    If we rewrite the above equation using $f^l(t,r)$, the $\mu=b$ component of Maxwell equation leads to
    \begin{equation}
        D_a\left[C(r)\left(D^a f^l(t,r) + \epsilon^{ab} r^2 j^l_b(t,r) \right)\right]=\frac{f^l(t,r)}{r^2}\ .
    \end{equation}
    $C(r)$ is given in \eqref{eq:scalar_mode_C(r)}. From Eq.~(\ref{scalareq}), we can reproduce $a^l_b(t,r)$ from $f^l(t,r)$ as
    \begin{equation}
    a^l_b(t,r) = C(r)[\epsilon_{ab} D^a f^l(t,r) - r^2 j_b]
    \end{equation}
    By taking the limit of $r\to \infty$ in this equation, we have Eq.~(\ref{eq:scalar_mode_a(f)}).

    Next, the $\mu=i$ component of the Maxwell equation is given by
    \begin{equation}
        \nabla_\nu\delta F^{\nu i}= \delta J^i\ .
    \end{equation}
    The right hand side will be 
    \begin{equation}
        \delta J^i
        =i \sum_l c_l \left(\Psi^*\psi^l-\Psi\psi^{*l}\right)\frac{1}{r^2}\hat{D}^i Y_{l0}(\theta)\ .
    \end{equation}
    The left hand side will be
    \begin{equation}
        \nabla_\nu\delta F^{\nu i}=-\frac{1}{r^2}  \sum_l c_l D^a a^l_a(t,r)\ \hat{D}^i Y_{l0}(\theta)\ .
    \end{equation}
    By comparing both sides, we get
    \begin{equation}
    \label{eq:Maxwell_i}
        D^a a^l_a(t,r) = -i\left(\Psi^*\psi^l(t,r)-\Psi\psi^{*l}(t,r)\right)\ .
    \end{equation}

    Finally, the equation of motion of the charged scalar field is 
    \begin{equation}
        \mathcal{D}^2\delta \Psi +\frac{2}{L^2}\delta \Psi=2i \delta A^\mu\mathcal{D}_\mu\Psi+i(\nabla_\mu\delta A^\mu)\Psi\ .
    \end{equation}
    The first term of left hand side leads to 
    \begin{equation}
        \mathcal{D}^2\delta \Psi
        =\sum_l c_l \left\{
        \mathcal{D}_a\mathcal{D}^a
        +\frac{2}{r}\left(\partial_a r\right)\mathcal{D}^a
        -\frac{l(l+1)}{r^2}
        \right\}\psi^l(t,r) Y_{l0}(\theta)\ ,
    \end{equation}
    and the second term of left hand side is
    \begin{equation}
        \frac{2}{L^2}\delta\Psi=\sum_l c_l \frac{2}{L^2}\psi^l(t,r) Y_{l0}(\theta)
    \end{equation}
    For the right hand side,
    \begin{align}
        \nonumber&
        2i\delta A^\mu\mathcal{D}_\mu\Psi+ i(\nabla_\mu\delta A^\mu)\Psi\\
        \nonumber&
        =\sum_l c_l \left[
        2ia^l_a(t,r)\mathcal{D}^a\Psi
        +i\left( D^a a^l_a(t,r)+\frac{2}{r}\left(\partial^a r\right)a^l_a(t,r)\right)\Psi
        \right]Y_{l0}(\theta)\\
        &
        =\left[
        2i C(r)\left(\epsilon_{ba}D^b f-r^2 j^l_a\right)\left(\mathcal{D}^a+\frac{\partial^a r}{r}\right)
        +\Psi^*\psi^l-\Psi\psi^{*l}
        \right]\Psi Y_{l0}(\theta)\ .
    \end{align}
    We used \eqref{eq:Maxwell_i} and \eqref{eq:scalar_mode_a(f)} to derive the third equality.
    Then, by comparing the both sides, we get \eqref{eq:scalar_mode_KG}:
    \begin{align}
        \nonumber&
        \left\{
        \mathcal{D}_a\mathcal{D}^a + \frac{2}{r}g^{ar}\mathcal{D}_a - \frac{l(l+1)}{r^2} + \frac{2}{L^2}
        \right\}\psi^l(t,r)\\
        &
        =\left\{
        2iC(r)\left(\epsilon^{ba}D_b f^l-r^2 (j^l)^a\right) \left(\mathcal{D}_a+\frac{\partial_a r}{r}\right)
        +\Psi^*\psi^l-\Psi\psi^{*l}\right\}\Psi\ .
    \end{align}

\section{The source and the response}
\label{appendix_response}
    We will derive an expression of the response \eqref{eq:vector_mode_response}, \eqref{eq:scalar_mode_response_t} and \eqref{eq:scalar_mode_response_i} in this section.

    The Maxwell action in Sch-AdS spacetime is given as
    \begin{equation}
        S=-\frac{1}{4} \int d^4x \sqrt{-g}F_{\mu\nu}F^{\mu\nu}\ ,
    \end{equation}
    where $g$ represents a determinant of the metric \eqref{eq:SAdS metric}.
    Let us calculate the electric current at the AdS boundary with respect to an infinitesimal change of a boundary value of a gauge field $A_\mu|_{r=\infty}$. Due to a change in $A_\mu|_{r=\infty}$, a gauge field in the bulk also varies as $A_\mu \rightarrow A_\mu + \Delta A_\mu$. The deviation of the Maxwell action is
    \begin{align}
        \Delta S
        \nonumber&=-\frac{1}{2} \int d^4x \sqrt{-g} F_{\mu\nu} F^{\mu\nu}
        =-\int d^4x \sqrt{-g} F^{\mu\nu}\nabla_\mu \Delta A_\nu\\
        \nonumber&=-\int \sqrt{-g} \nabla_\mu\left( F^{\mu\nu}\Delta A_\nu \right)
        =-\int d^3x \sqrt{-h} n_\mu F^{\mu\nu}\Delta A_\nu|_{r=\infty}\ .
    \end{align}
    Here, we introduced an induced metric on the constant $r$ hypersurface: $h_{IJ}$ ($\{I,J\}=\{t,\theta,\phi\}$). $h$ is its determinant and $n_\mu$ is its unit normal vector.
    We used the Maxwell equation $\nabla_\mu F^{\mu\nu}=0$ at the third equality. 
    Near the AdS boundary, we get $h_{IJ}dx^I dx^J \simeq r^2(-dt^2 + d\Omega^2)$ and $n_\mu \simeq r^{-1} (dr)_\mu$. 
    We define the boundary electric current as
    \begin{equation}
        \langle J^I \rangle = \frac{1}{\sqrt{-\tilde{h}}}\frac{\Delta S}{\Delta A_I}=-r^2 F^{rI}|_{r=\infty}\ ,
    \end{equation}
    where $\tilde{h}$ is the determinant of rescaled metric: $\tilde{h}_{IJ} \simeq -dt^2 + d\Omega^2\ \ (r\to\infty)$. The boundary electric current satisfies the conservation law automatically:
    \begin{equation}
        \nabla_I \langle J^I \rangle  = -r^2 \nabla_I F^{rI}=0\ .
    \end{equation}

    For the vector mode, the electromagnetic tensors at the AdS boundary are
    \begin{equation}
        F^{rt}|_{r=\infty} = 0\ ,
        \ \ \ 
        F^{ri}|_{r=\infty} \simeq -\frac{1}{r^2} e^{-i\omega t}\sum_l c_l \alpha^{l(1)} (Y_{l0}(\theta))^i\ ,
    \end{equation}
    where we used \eqref{eq:vector_mode_asmp} for $F^{ri}$. This gives the electric current in Eq.~(\ref{eq:vector_mode_response}).

    For the scalar mode, we have
    \begin{align}
        F^{rt}&= \frac{1}{r^2} e^{-i\omega t}\sum_l c_l f^{l(0)} Y_{l0}(\theta)\ ,\\
        F^{ri}&= -\frac{1}{r^2}\, i\omega e^{-i\omega t}\sum_l \frac{1}{l(l+1)} c_l f^{l(0)} \hat{D}^i Y_{l0}(\theta)\ ,
    \end{align}
    where we have used \eqref{eq:scalar_mode_a(f)} to get $F^{ri}$. This gives the electric current in Eqs.~\eqref{eq:scalar_mode_response_t} and \eqref{eq:scalar_mode_response_i}.

\section{WKB analysis}
\label{appendix_WKB}

    In the normal phase of the holographic superconductor, Maxwell perturbation equations for vector and scalar modes are identical.
    They are uniformly written in the Schr\"{o}dinger form as
    \begin{equation}
        \left[-\frac{d^2}{dr_\ast^2}+U(r)\right]\chi=0\ ,\quad U(r)=-\omega^2+l(l+1)v(r)\ ,
        \label{vecmode}
    \end{equation}
    where $\chi=\alpha^l$ and $\chi=f^l$ for the vector and scalar modes, respectively.
    The effective potential $v(r)$ is defined in Eq.~(\ref{geodesic}).
    The asymptotic solution near the infinity is
    \begin{equation}
        \chi(r)=\chi_0 + \frac{\chi_1}{r}+ \mathcal{O}\left(\frac{1}{r^2}\right) = \chi_0 - \chi_1 r_\ast + \mathcal{O}(r_\ast^2)\ .
    \end{equation}
    In the followings, we will determine the ratio of the source and response, $\chi_1/\chi_0$, by WKB approximation. 

\begin{figure}
        \begin{center}
        \includegraphics[scale=0.5]{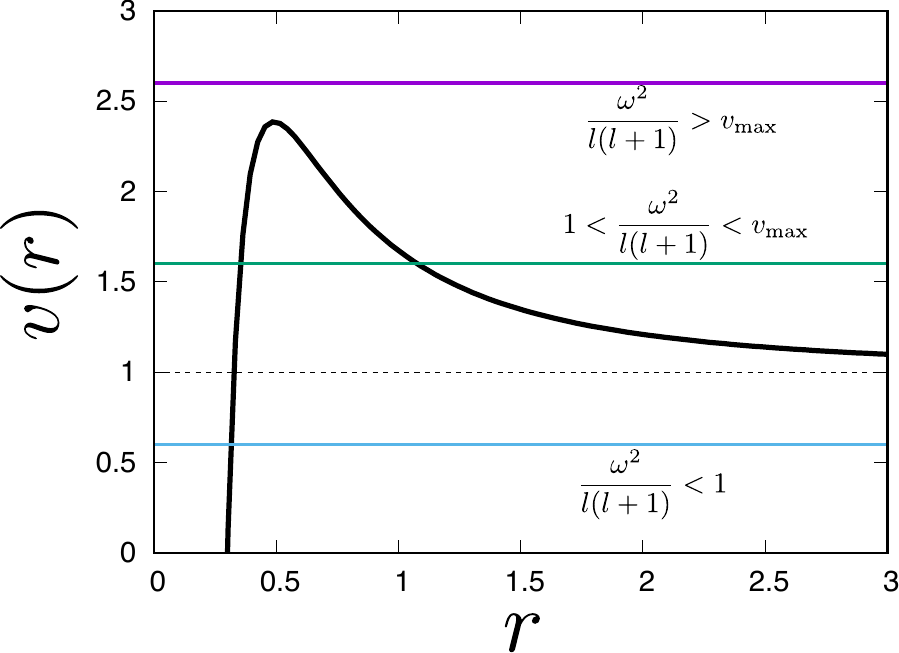}
        \end{center}
        \caption{Effective potential for Maxwell perturbation equations for $r_h=0.3$\ .
        }
        \label{Veff}
    \end{figure}

    Fig.~\ref{Veff} shows the effective potential $v(r)$ for $r_h=0.3$.
    The potential has the maximum value $v_\textrm{max}$ defined in Eq.~(\ref{vmax}).
    For the WKB analysis, we need to consider three cases depending on the number of turning points:
    $l(l+1) v_\textrm{max} <  \omega^2$, $l(l+1) < \omega^2 < l(l+1) v_\textrm{max}$, and  $\omega^2 < l(l+1)$.

\subsection{\texorpdfstring{$l(l+1) v_\textrm{max} <  \omega^2$}{Lg}}

    Firstly, we consider the case of $l(l+1) v_\textrm{max} < \omega^2$. There is no turning point in this case and the WKB solution is simply given by
    \begin{equation}
        \chi(r_\ast) = \frac{1}{(-U)^{1/4}} \exp\left[-i\int_0^{r_\ast} dr_\ast  \sqrt{-U}\right]\ ,
        \label{WKBa0}
    \end{equation}
    where we took in-going mode at the horizon. Near infinity $r_\ast\sim 0$, this solution behaves as
    \begin{equation}
        \chi(r_\ast)\simeq \frac{1}{\sqrt{k}} \exp(-ikr_\ast) \simeq \frac{1}{\sqrt{k}}(1-ikr_\ast)\ ,
    \end{equation}
    where $k^2=\omega^2-l(l+1)$. Therefore, we have
    \begin{equation}
        \frac{\chi_1}{\chi_0}=ik\ .
    \end{equation}

\subsection{\texorpdfstring{$\omega^2<l(l+1)$}{Lg}}

    Secondly, let us consider the case of $\omega^2<l(l+1)$. There is single turning point in this case.
    We denote the turning point as $r_\ast=a$, i.e., $U(a)=0$. For $r_\ast<a$, the WKB solution is 
    \begin{equation}
        \chi(r_\ast<a) = \frac{1}{(-U)^{1/4}} \exp\left[-i\int_a^{r_\ast} dr_\ast  \sqrt{-U}\right]\ .
        \label{WKBa}
    \end{equation}
    This is essentially same solution as Eq.~(\ref{WKBa0}), but we took $r_\ast=a$ as the lower bound of the integration for later convenience.
    For $r_\ast>a$, we obtain
    \begin{equation}
        \begin{split}
         \chi(r_\ast>a) &= \frac{e^{-i\pi/4}}{U^{1/4}}\left\{\exp\left[\int_a^{r_\ast} dr_\ast  \sqrt{U}\right]
        +\frac{i}{2}\exp\left[-\int_a^{r_\ast} dr_\ast  \sqrt{U}\right]
        \right\}\\
        &= \frac{e^{-i\pi/4+\Gamma_0/2}}{U^{1/4}}\left\{\exp\left[\int_0^{r_\ast} dr_\ast  \sqrt{U}\right]
        +\frac{i}{2}e^{-\Gamma_0}\exp\left[-\int_0^{r_\ast} dr_\ast  \sqrt{U}\right]\right\}\ .
        \end{split}
        \label{WKBab}
    \end{equation}
    At the last equality, we used $\int_a^{r_\ast}=\int_a^0+\int_0^{r_\ast}$ and defined
    \begin{equation}
        \Gamma_0 \equiv 2\int_a^0 dr_\ast  \sqrt{U}\ .
    \end{equation}
    Near the infinity, the solution behaves as
    \begin{equation}
        \begin{split}
         \chi(r_\ast)&\simeq \frac{e^{-i\pi/4+\Gamma_0/2}}{\sqrt{\kappa}}\left\{e^{\kappa r_\ast}
        +\frac{i}{2}e^{-\Gamma_0}e^{-\kappa r_\ast}\right\}\\
        &\simeq\frac{e^{-i\pi/4+\Gamma_0/2}}{\sqrt{\kappa}}\left\{1+\frac{i}{2}e^{-\Gamma_0}+\kappa (1-\frac{i}{2}e^{-\Gamma_0}) r_\ast \right\}\ ,
        \end{split}
    \end{equation}
    where $\kappa^2=l(l+1)-\omega^2$. 
    Thus, we have
    \begin{equation}
        \frac{\chi_1}{\chi_0}= -\kappa \frac{2-ie^{-\Gamma_0}}{2+ie^{-\Gamma_0}}
    \end{equation}

\subsection{\texorpdfstring{$l(l+1) < \omega^2 < l(l+1) v_\textrm{max}$}{Lg}}

    Finally, we consider $l(l+1) < \omega^2 < l(l+1) v_\textrm{max}$, where there are two turning points $r_\ast=a,b$ ($a<b$).
    For $r_\ast<a$, the WKB solution is same as Eq.~(\ref{WKBa}).
    For $a<r_\ast <b$, the solution is also same as Eq.~(\ref{WKBab}) but it is convenient to rewrite it as
    \begin{equation}
     \chi(a<r_\ast < b) = \frac{e^{-i\pi/4+\Gamma/2}}{U^{1/4}}\left\{\exp\left[\int_b^{r_\ast} dr_\ast  \sqrt{U}\right]
    +\frac{i}{2}e^{-\Gamma}\exp\left[-\int_b^{r_\ast} dr_\ast  \sqrt{U}\right]\right\}\ ,
    \end{equation}
    where  we use $\int_a^{r_\ast}=\int_a^b+\int_b^{r_\ast}$ and define
    \begin{equation}
     \Gamma \equiv 2\int_a^b dr_\ast  \sqrt{U}\ .
    \end{equation}
    For $b<r_\ast$, the WKB solution is written as
    \begin{equation}
     \chi(r_\ast >b) = \frac{e^{\Gamma/2}}{(-U)^{1/4}}\left\{c_+\exp\left[i\int_b^{r_\ast} dr_\ast  \sqrt{-U}\right]
    +c_- \exp\left[-i\int_b^{r_\ast} dr_\ast  \sqrt{-U}\right]
    \right\}\ .
    \end{equation}
    where 
    \begin{equation}
    c_+=-i\, \left(1-\frac{e^{-\Gamma}}{4}\right)\ ,\quad 
    c_-=  e^{-\Gamma}\left(1+\frac{e^{-\Gamma}}{4}\right)\ .
    \end{equation}
    Again, we determined $c_\pm$ using standard connection formulae of WKB.
    From $\int_b^{r_\ast}=\int_b^0+\int_0^{r_\ast}$, above expression becomes
    \begin{equation}
    \chi(r_\ast >b) = \frac{e^{\Gamma/2+i\delta/2}}{(-U)^{1/4}}\left\{c_+ \exp\left[i\int_0^{r_\ast} dr_\ast  \sqrt{-U}\right]
    +c_-e^{-i\delta} \exp\left[-i\int_0^{r_\ast} dr_\ast  \sqrt{-U}\right]
    \right\}\ ,
    \end{equation}
    where
    \begin{equation}
    \delta \equiv 2\int_b^0 dr_\ast  \sqrt{-U}\ .
    \end{equation}
    Asymptotic solution near infinity is
    \begin{equation}
    \chi(r_\ast) \simeq \frac{e^{\Gamma/2+i\delta/2}}{k^{1/2}}\left\{(c_++c_-e^{-i\delta})
    +ik(c_+-c_-e^{-i\delta})r_\ast
    \right\}
    \end{equation}
    Therefore, we obtain
    \begin{equation}
    \begin{split}
     \frac{\chi_1}{\chi_0}&=-ik\,\frac{c_+-c_- e^{-i\delta}}{c_++c_-e^{-i\delta}}\\
    &=k\, \frac{e^{-\Gamma} \tan(\delta/2+\pi/4) +4i}{e^{-\Gamma}- 4i \tan(\delta/2+\pi/4)}\ .
    \end{split}
    \end{equation}

\subsection{WKB and full numerical solutions}
\label{WKBfull}

\begin{figure}
        \centering
        \subfigure
         {\includegraphics[scale=0.4]{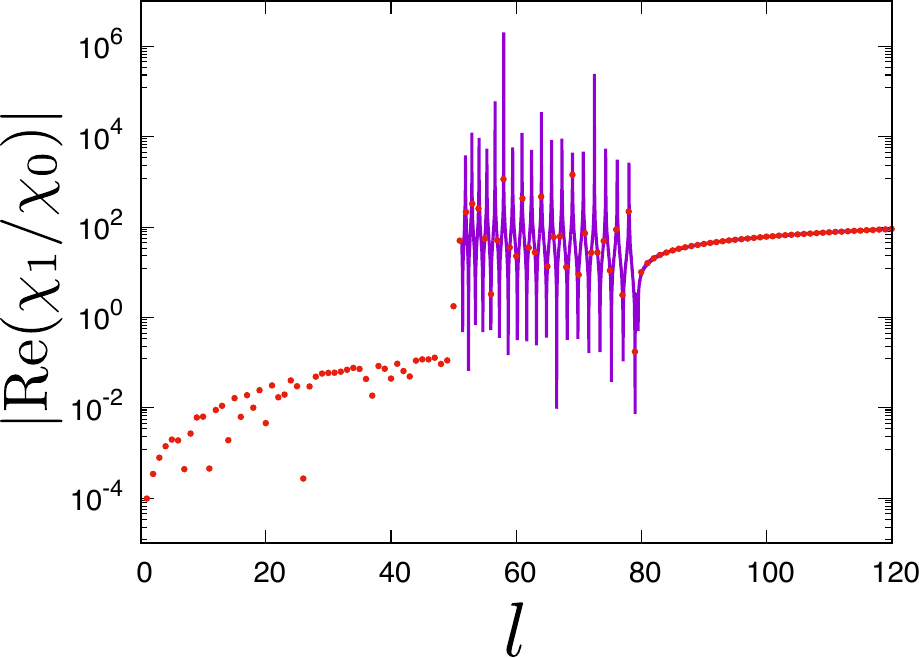}\label{ReWKB}
          }
          \subfigure
         {\includegraphics[scale=0.4]{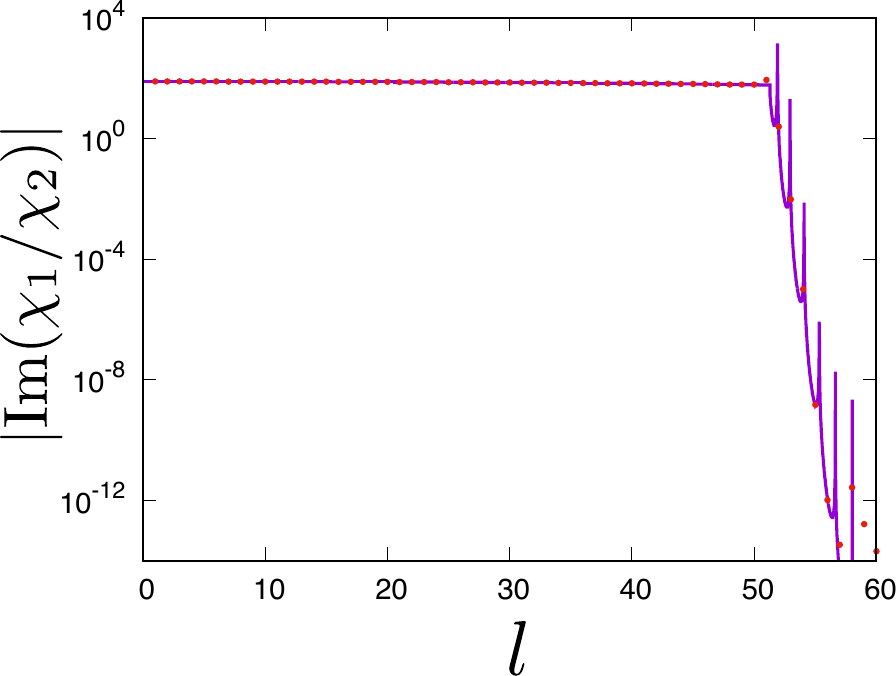}\label{ImWKB}
          }
         \caption{
        Comparison of $\chi_1/\chi_0$ for $r_h=0.3$
        between WKB approximation (purple curve) and full numerical calculation (red dots).
        In the WKB analysis, the quantum number of the spherical harmonics $l$ is regarded as a continuous parameter for visibility.
        }
        \label{ReImWKB}
    \end{figure}

    Fig.~\ref{ReImWKB} shows $\chi_1/\chi_0$ obtained by the WKB approximation and full numerical calculation.
    The background is the Schwarzschild-AdS with $r_h=0.3$.
    In the WKB analysis, we formally regard $l$ as a continuous parameter. There is a good agreement between them.
    When the two turning points $r_\ast=a,b$ are separated enough, the tunneling probability $e^{-\Gamma}$ is highly suppressed. 
    Then, we have
    \begin{equation}
        \textrm{Re}\left(\frac{\chi_1}{\chi_0}\right)\simeq -\frac{k}{\tan(\delta/2+\pi/4)}\ ,\quad 
        \textrm{Im}\left(\frac{\chi_1}{\chi_0}\right)\simeq \frac{k e^{-\Gamma}}{4 \sin^2 (\delta/2+\pi/4)}\ .
        \label{ReIm}
    \end{equation}
    When $\delta/2+\pi/4 \simeq \pi m$ $(m\in \bm{Z})$, real and imaginary parts of $\chi_1/\chi_0$ take large values 
    and this is the origin of spikes found in Fig.~\ref{ReImWKB}.
    The modes with $\delta/2+\pi/4 \simeq \pi m$ correspond to 
    ``normal modes'' trapped in the outside of the potential $b<r_\ast<0$. 
    (To be precise, they should be regarded as quasinormal modes with tiny damping factors because of the 
    small tunneling probability $e^{-\Gamma}$.) 
    Spikes found in the response function is caused by the reflection of the bulk wave at the time like boundary and this phenomena is characteristic to the asymptotically AdS spacetime.
    In general, $\delta/2+\pi/4 = \pi m$ is not exactly satisfied since $l$ is an integer and $\delta$ takes discrete values.
    However, depending on the parameters $\omega$ and $r_h$, $\delta/2+\pi/4$ ``accidentally'' has a  value close to $\pi m$ and then $\chi_1/\chi_0$ becomes large.
    As discussed in section~\ref{sec:Review}, if the response function has a spiky point at $l=l_0$, we find the ring in the image at $\sin\theta_s = l_0/\omega$
    Therefore, if we make the image from the response function, 
    it can be sensitive on the parameters and this is actually found in our previous work~\cite{Hashimoto:2018okj,Hashimoto:2019jmw}. 
    This is the reason why we proposed the prescription to take only imaginary part of the response for imaging.

\subsection{Analytical calculation of the radius of the photon ring}
\label{ImagingfromIm}

    As we can see in Eq.~(\ref{ReIm}) or Fig.~\ref{ImWKB}, 
    the imaginary part of the response is exponentially suppresed in the region of $l(l+1) v_\textrm{max} >  \omega^2$ because of the tiny tunneling probability $e^{-\Gamma}$.
    Thus, we would be able to approximate the imaginary part of the respose function as
    \begin{equation}
        \textrm{Im}\left(\frac{\chi_1}{\chi_0}\right) \simeq  
        \begin{cases}
         \sqrt{\omega^2-l(l+1)} & (l <  \omega/\sqrt{v_\text{max}})\\
         0 & (\textrm{otherwise})
        \end{cases}
        \ .
        \label{ImRes}
    \end{equation}
    Let us consider the image of the black hole constructed from this imaginary part of the response function.
    For simplicity, we consider the case of $\sigma=0$, i.e., the external electromagnetic field is given by the delta function. 

    Then, we can prove
    \begin{equation}
     \sum_{l=0}^n c_l Y_{l0}(\theta) \simeq  \frac{1}{2} c_n Y_{n0}(\theta)\ ,\quad (n\gg1,\quad \theta \ll 1)\ ,
    \label{sumcY}
    \end{equation}
    where $c_l=(-1)^l (2l+1)^{1/2}/(4\pi)^{1/2}$, which is equivalent to  Eq.~(\ref{eq:vector_mode_cl}) with $\sigma=0$.
    The proof is the follows:
    \begin{equation}
        \begin{split}
         \sum_{l=0}^n c_l Y_{l0}(\theta)
        &=\frac{1}{4\pi}\sum_{l=0}^n (-1)^l (2l+1) P_l(\cos \theta)\\
        &=\frac{1}{4\pi}\sum_{l=0}^n  (2l+1) P_l(-1) P_l(\cos \theta)\\
        &=\frac{1}{4\pi}\frac{n+1}{1+\cos\theta}\{P_{n+1}(\cos\theta)P_n(-1)-P_{n}(\cos\theta)P_{n+1}(-1)\}\\
        &=\frac{1}{4\cos^2\theta/2}\left\{
        \frac{n+1}{n+1/2} c_n Y_{n0}(\theta)- \frac{n+1}{n+3/2} c_{n+1}Y_{n+1,0}(\theta)
        \right\}
        \end{split}
    \end{equation}
    where $P_l$ is the Legendre polynomial. At the second equality, we used $P_l(-1)=(-1)^l$. 
    At the third equality, we used the finite-sum-formula for the Legendre polynomial~\cite{Legendre}:
\begin{equation}
    \sum_{l=0}^n (2l+1)P_l(z_1)P_l(z_2)=\frac{n+1}{z_1-z_2}\{P_{n+1}(z_1)P_{n}(z_2)-P_{n}(z_1)P_{n+1}(z_2)\}\ .
\end{equation}
    For $\theta\ll 1$, the spherical harmonics is approximated as 
    \begin{equation}
     Y_{l0}(\theta)\simeq \sqrt{\frac{l+1/2}{2\pi}} J_0((l+1/2)\theta)\ ,
    \label{Yapprox}
    \end{equation}
    where $J_m$ is the Bessel function. Since above function is continuous function of $l$, we have $Y_{n0}(\theta) \simeq Y_{n+1,0}(\theta)$ for $n\gg1$ and $\theta \ll 1$.
    We also obtain $c_{n+1}\simeq - c_n$ for $n\gg1$. Then it leads to Eq.~(\ref{sumcY}).

    From Eq.~(\ref{ImRes}), the imaginary part of the response function in the real space is given by
    \begin{equation}
    \begin{split}
    \textrm{Im}\langle\mathcal{O}(\theta)\rangle &\simeq \sum_{l=0}^n \sqrt{\omega^2-l(l+1)} c_l Y_{l0}(\theta) = \sqrt{\omega^2+\hat{D}^2} \sum_{l=0}^n c_l Y_{l0}(\theta)\\
    &\simeq \frac{1}{2}\sqrt{\omega^2-n(n+1)} c_n Y_{n0}(\theta)\ ,
    \end{split} 
    \end{equation}
    where we define $n=\omega/\sqrt{v_\textrm{max}}$. This is proportional to the single spherical harmonics $Y_{n0}$. 
    Then, as shown in section.\ref{sec:Review}, we find the ring at
    \begin{equation}
     \sin\theta_S = \frac{n}{\omega} = \frac{1}{\sqrt{v_\textrm{max}}}\ .
\label{WKBthetaS}
    \end{equation}
    This coincides with the result of the geodesic approximation~(\ref{Estnradius}).

\bibliographystyle{JHEP}
\bibliography{ref.bib}

\end{document}